\documentclass[journal]{IEEEtran}

\usepackage{amsfonts}
\usepackage{bm}
\usepackage{amsmath,amsfonts,amsthm,amssymb}
\usepackage{setspace}
\usepackage{Tabbing}
\usepackage{fancyhdr}
\usepackage{lastpage}
\usepackage{extramarks}
\usepackage{chngpage}
\usepackage[normalem]{ulem}

\usepackage{array}
\usepackage[caption=false,font=normalsize,labelfont=sf,textfont=sf]{subfig}
\usepackage{textcomp}
\usepackage{stfloats}
\usepackage{url}
\usepackage{verbatim}
\usepackage[pdf]{pstricks}
\usepackage{graphicx}
\usepackage{textcomp}
\usepackage{xcolor, cite}
\usepackage{url}
\usepackage{multirow}
\usepackage{booktabs}
\usepackage{multirow}
\usepackage{footnote}
\usepackage{threeparttable}
\usepackage{nccmath}
\usepackage[caption=false,font=normalsize,labelfont=sf,textfont=sf]{subfig}
\usepackage{bbm}

\usepackage{algorithm2e}
\makeatletter
\newcommand{\nosemic}{\renewcommand{\@endalgocfline}{\relax}}
\newcommand{\dosemic}{\renewcommand{\@endalgocfline}{\algocf@endline}}
\let\oldnl\nl
\newcommand{\nonl}{\renewcommand{\nl}{\let\nl\oldnl}}
\makeatother

\usepackage{soul,color}
\usepackage{epsfig}
\usepackage{graphicx,float,wrapfig, enumerate}
\usepackage{dsfont}
\usepackage{longtable}
\usepackage{wrapfig}
\usepackage{stfloats}
\usepackage{cite}
\usepackage{graphicx}
\usepackage{array}
\usepackage{multirow}{\tiny }
\usepackage{multicol}
\usepackage{colortbl}
\usepackage{tabularx}
\usepackage{mdwmath}
\usepackage{mdwtab}
\usepackage{color}
\usepackage{verbatim}
\usepackage{amsmath}
\usepackage{tikz}
\usetikzlibrary{calc}
\usepackage{flowchart}
\usetikzlibrary{shapes.geometric, arrows}
\usepackage{overpic}
\usepackage{epstopdf}
\usepackage{url}
\usepackage{booktabs}

\usepackage{makecell}
\usepackage{textcomp}
\usepackage{bbm}

\newtheorem{lemma}{Lemma}

\newtheorem{remark}{Remark}

\usepackage{threeparttable}

\DeclareMathOperator*{\argmin}{arg\,min}

\usepackage{color}
\makeatletter %
\let\myorg@bibitem\bibitem
\def\bibitem#1#2\par{%
	\@ifundefined{bibitem@#1}{%
		\myorg@bibitem{#1}#2\par
	}{%
		\begingroup
		\color{\csname bibitem@#1\endcsname}%
		\myorg@bibitem{#1}#2\par
		\endgroup
	}%
}

\makeatother %

\allowdisplaybreaks

\usepackage{footnote}
\makesavenoteenv{tabular}
\makesavenoteenv{table}

\newcommand{\grn}[1]{\textcolor{black}{#1}}
\newcommand{\cg}[1]{\textcolor{black}{#1}}
\newcommand{\jq}[1]{{\color{black}#1}}

\begin{document}

\title{
Federated Aggregation of Demand Flexibility
	}

\author{
Yifan~Dong,~\IEEEmembership{Graduate Student Member,~IEEE,}
Ge~Chen,~\IEEEmembership{Member,~IEEE,}
and Junjie~Qin,~\IEEEmembership{Member,~IEEE}
 \vspace{-4mm}
\thanks{
This paper is supported in part by the U.S.  National Science Foundation under Grant No. EEC-1941524 and under Grant No. ECCS-2339803. 



G. Chen is with the School of Advanced Engineering, Great Bay University, Dongguan, China.
Y. Dong, and J. Qin are with the Elmore Family School of Electrical and Computer Engineering, Purdue University, West Lafayette, IN.}
}

\maketitle

\begin{abstract}

This paper proposes a federated framework for demand flexibility aggregation to support grid operations. Unlike existing  geometric  methods that rely on a static, pre-defined base set as the geometric template for aggregation, our framework establishes a true federated process by enabling the collaborative optimization of this base set without requiring the participants sharing sensitive data with the aggregator. Specifically, we first formulate the base set optimization problem as a bilevel program. Using optimal solution functions, we then reformulate the bilevel program into a single-level, unconstrained learning task. By exploiting the decomposable structure of the overall gradient, we further design a decentralized gradient-based algorithm to solve this learning task. The entire framework, encompassing base set optimization, aggregation, and disaggregation, operates by design without exchanging raw user data. Numerical results demonstrate that our proposed framework unlocks substantially more flexibility than the approaches with static base sets, thus providing a promising framework for efficient and privacy-enhanced approaches to coordinate demand flexibility at scale. 

\end{abstract}

\begin{IEEEkeywords}
 Demand flexibility, flexibility aggregation, federated algorithms,  geometric methods.
\end{IEEEkeywords}

\section{Introduction} \label{sec_intro}

The large-scale aggregation of demand flexibility offers a highly effective pathway for maintaining grid operation security against the growing intermittency of renewable sources and surging demands from new loads \cite{denholm2021challenges, 10965352}. To harness the collective flexibility potential of vast populations of demand-side resources (DSRs) like electric vehicles (EVs) and thermostatically controlled loads (TCLs), 
developing scalable and data-driven aggregation mechanisms is paramount \cite{10700765}.
However, a fundamental challenge arises: achieving high aggregation performance typically requires access to sensitive information of individual DSRs, \cg{such as the travel information for EVs and occupancy information for building TCLs}, which directly conflicts with the imperative of user privacy \cite{10681446}. 

This critical need for privacy-preserving yet high-performing aggregation methods for demand flexibility has spurred the search for new decentralized architectures. Federated learning, in particular, offers a powerful conceptual blueprint. It enables participants to collaboratively build a shared model by exchanging only intermediate parameters while keeping raw data local; analogously (though not for machine learning model training), this motivates a framework with similar information-exchange patterns and privacy properties for  aggregation of demand flexibility to resolve the aforementioned conflict \cite{9478223}.

In the context of flexibility aggregation, decentralized geometric methods \cite{8063901, 7864461} represent the closest existing approach to this federated scheme: an aggregator defines and distributes a shared model, a \emph{base set}, to a population of DSRs. Each DSR then performs local computation, solving an optimization problem to find its best affine transformation of this base set, and sends only the resulting low-dimensional parameters back to the aggregator. This process, by design, avoids the sharing of sensitive user data, thus offering clear advantages in scalability and privacy preservation.

Despite these structural advantages, this ``proto-federated" approach diverges from a true federated process. Its core limitation is that the shared model, the base set, is treated as a static, heuristic template, rather than as an optimizable variable to be collaboratively updated based on DSRs' local data \cite{8063901}. This static-template design treats DSRs as passive participants instead of active learners who can jointly update the shared model tailored to the population's unique characteristics. This absence of a collaborative optimization process may create an unavoidable performance bottleneck for demand flexibility aggregation.

A natural path to resolve this bottleneck is to treat the base set as an optimizable decision variable. However, this straightforward idea introduces methodological and architectural challenges. First, at the \grn{user} level, the subproblem of finding the optimal affine transformation becomes non-convex. This non-convexity arises because treating the base set as a decision variable, alongside the transformation parameters, introduces bilinear terms in the set-containment constraint \cite{9029363}.
Second, at the system level, optimizing a shared base set inherently couples all participants. \cg{This coupling traditionally leads to} a centralized formulation that requires the aggregator to collect each \grn{user’s} private information, thereby violating the foundational principles of a federated architecture. 
Therefore, \cg{\emph{designing a federated method for base set optimization without sharing sensitive user data with an aggregator remains an open problem.}} 

To address this challenge, this paper proposes a \cg{federated framework for demand flexibility aggregation}. Our framework enables the collaborative optimization of the base set by first reformulating the corresponding non-convex problem into an unconstrained learning task. We then solve this task using a decentralized gradient-based algorithm that leverages the problem's inherently decomposable structure. Finally, we complete the framework by integrating existing protocols for the aggregation and disaggregation workflow. The main contributions of this work are threefold:

\begin{enumerate}
	\item We propose a novel federated framework that, for the first time, unifies the processes of base set optimization, aggregation, and disaggregation into a decentralized architecture. This design ensures that all sensitive user data remains strictly local, with only anonymized model updates exchanged throughout the entire workflow. It simultaneously unlocks substantially better aggregation performance by incorporating an optimizable base set.
	
	\item We develop an efficient method for federated optimization 
    of the base set. Our method provides a tractable solution pathway by reformulating the intractable base set optimization problem, which is bilevel and non-convex, into a single-level unconstrained learning task using optimal solution functions. We then design a decentralized gradient algorithm to solve this task. Since this algorithm relies on the inherently decomposable structure of the gradient, it provably incurs no optimality loss compared to a centralized implementation.
	
	\item We demonstrate that existing decentralized aggregation and disaggregation protocols can be seamlessly integrated into our federated framework. This integration completes the architecture, providing a ready-to-deploy system for demand flexibility aggregation.
\end{enumerate}


The remaining parts are organized as follows: Section \ref{sec:literature} reviews related literature. Section \ref{sec:framework} outlines the \cg{federated pipeline for base set optimization, aggregation and disaggregation}, and formulates the base set optimization. Section \ref{sec:method} \cg{introduces the federated method for base set optimization.} 
Section \ref{sec:protocols} details the protocols for federated aggregation and disaggregation. Section \ref{sec:simulation} presents case studies, and  Section \ref{sec:conclusion} concludes the paper.

\section{Related Literature}\label{sec:literature}

Existing literature on aggregating demand flexibility\footnote{Beyond explicitly aggregating the demand flexibility sets, there is a broader literature on coordinating DSRs (cf. \cite{7445245, 10073567} and references therein).} can be broadly classified into three categories: boundary aggregation, optimization-based and geometric methods. 

\subsubsection{Boundary aggregation methods}
The studies in this category model each DSR as a virtual battery by identifying the upper and lower bounds of realizable charging power and cumulative energy \cite{10878458, 11045802}. These individual boundaries are further aggregated to form a larger virtual battery for aggregate flexibility characterization. For example, \cg{Zhang et al. \cite{7463483} and Brinkel et al.} \cite{BRINKEL2023100297}
first derive the realizable upper and lower power-and-energy envelopes for each individual DSR and then sum these envelopes to obtain aggregate flexibility bounds. However, this approach typically assumes homogeneous DSR behavior (e.g., similar plug-in durations) and the resulting power-and-energy envelopes may be outer approximations, in which many trajectories cannot actually be scheduled for individual DSRs \cite{9851563}.

\subsubsection{Optimization-based methods}
To address this issue, the second category, optimization-based methods, characterizes the aggregated \grn{demand} flexibility by formulating and solving a two-stage robust optimization problem to identify the maximum-volume box or ellipsoid \cite{10124221, 10347534}. The first stage optimizes the boundaries of the chosen geometric shape to maximize its volume, while the second stage verifies whether each point within this shape can be realized by adjusting individual power profiles of DSRs \cite{9482808, 10286155}. Although this method can guarantee the realizability of the identified aggregate flexibility, it is usually computationally intensive, as it needs to repeatedly solve  mixed-integer problems \cite{9383807}. Furthermore, it  requires full knowledge of each DSR's charging behavior, raising privacy concerns for users.

\subsubsection{Geometric methods}
Geometric methods have emerged as computationally efficient tools for approximating aggregate \grn{demand} flexibility, 
with their decentralized structure offering a potential way to address the privacy issue.
These methods first predetermine a geometric template, known as the \emph{base set}, and then share it among all DSRs to identify the maximum-volume linear transformations of the base set within the flexibility sets of individual DSRs \cite{8063901, 7864461}. Since all these inner approximations are based on the geometric template, their Minkowski sum can be efficiently computed, resulting in a larger set to characterize the aggregate flexibility  \cite{10490133}. In this process, each \grn{user} independently and locally solves a linear program to obtain optimal linear transformations without sharing its charging behavior information with the aggregator. Therefore, these methods can avoid the sharing of sensitive user data. However, the performance of geometric methods highly depends on the choice of base set. An improper selection can lead to overly conservative results, shrinking the aggregate flexibility. Some studies, such as \cite{10490133}, mitigate this by constructing the base set using the averaged parameters of all DSRs. \cg{However, this is heuristic and may still produce conservative outcomes.} 
More importantly, this reintroduces privacy concerns by requiring access to data from every \grn{user} to calculate the average.

\section{\cg{Federated Framework for Demand Flexibility Aggregation}}\label{sec:framework}
\cg{This section introduces our proposed federated framework for demand flexibility aggregation. We begin with a high-level overview of the framework, where the end-to-end pipeline for base set optimization, aggregation and disaggregation is established.}
We then present the mathematical preliminaries for flexibility aggregation. Finally, we formulate the core technical challenge, which is base set optimization. 


\subsection{Overview of the Framework}

\begin{figure}[ht]
	\vspace{-3mm}
	\centering	{\includegraphics[width=0.98\columnwidth]{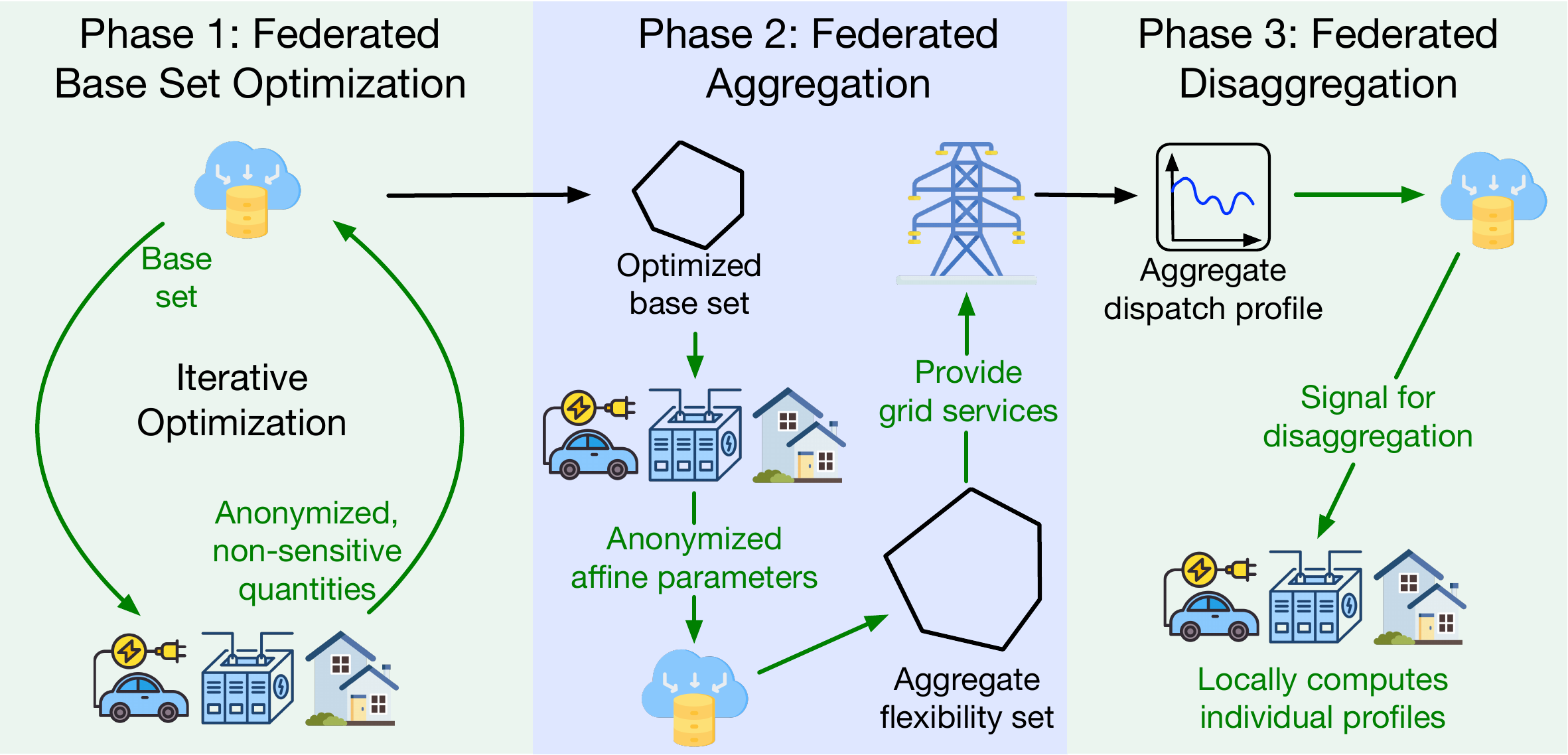}}
	\vspace{-3mm}
	\caption{\cg{The federated framework for demand flexibility aggregation}.}
	\vspace{-3mm}
	\label{fig_framework}
\end{figure}

Our proposed framework operates as a three-phase process, as illustrated in Fig. \ref{fig_framework}. This architecture is designed to enable the aggregation of flexibility from a large population of DSRs without sharing sensitive user data.

\begin{itemize}
	\item \textbf{Phase 1: Federated Base Set Optimization.} This is the core innovation of our framework. The aggregator and all participating DSRs engage in an iterative, federated optimization process to collaboratively learn the optimal shared geometric template (the \emph{base set}). This phase ensures that the model is tailored to the specific characteristics of the current resource population. The detailed methodology for this phase is the central topic of Section IV.
	
	\item \textbf{Phase 2: Federated Aggregation.} Once the optimal base set is learned (or if a pre-defined one is used in a static setting), the aggregator initiates an aggregation process. Each DSR locally computes its optimal affine transformation parameters based on the shared base set and anonymously returns these non-sensitive updates to the aggregator, which then constructs the aggregate flexibility set.
	
	\item \textbf{Phase 3: Federated Disaggregation.} 
	\cg{The aggregate flexibility is offered to the system operator for the provision of grid services. Upon acceptance, the operator dispatches this flexibility by issuing a corresponding aggregate profile to the aggregator.} 
The aggregator then broadcasts a corresponding signal back to the DSRs, allowing each participant to locally and privately compute its own feasible dispatch profile.
\end{itemize}

\subsection{Preliminaries for Flexibility Aggregation}
We establish the modeling of demand flexibility using EVs as examples for DSRs. Nevertheless, it is crucial to highlight the broader applicability of the modeling framework to other DSRs with temporal energy/demand shifting flexibility, such as  energy storage systems and TCLs.

\subsubsection{Individual flexibility set}
We consider a finite-horizon discrete-time model with time slots indexed by $t\in \mathcal T:=\{1, \dots, T\}$, and there are $N$ EVs indexed by $\mathcal N := \{1, \dots, N\}$. 
We denote the net charging power and cumulative charging energy as $\mathbf u_i \!:=\! \big[u_i(t)\big]_{\!t\in \mathcal T}$ and $\mathbf x_i \!:=\! \big[x_i(t)\big]_{\!t\in \mathcal T}$, respectively. Given a charging profile $\mathbf u_i$, the cumulative energy stored in the EV battery can be expressed as \cite{10121812}:\footnote{
We adopt a lossless charging model to ensure that individual EV flexibility sets admit a convex polyhedral representation (H-representation), which is the key structural requirement of the proposed framework. The framework directly applies to one-way charging scenarios, where $u_i(t)\ge 0$ for all $t\in\mathcal T$, by incorporating charging inefficiencies as constant scaling factors in the constraint matrices.
For general bidirectional lossy charging models, convex polyhedral inner approximations of the resulting nonconvex flexibility sets, admitting an H-representation, have been established in existing literature \cite{9527324,10384282}. Consequently, the proposed framework remains applicable when such approximations are employed.
}
\begin{equation}\label{eq:power_energy}
	\abovedisplayskip=4pt
	\belowdisplayskip=4pt
	x_i(t) = x_i(t-1) + u_i(t) \cdot \Delta t,  \quad t\in \mathcal T,\ i\in\mathcal N,
\end{equation}
where $\Delta t$ is the time interval and $x_i(0) := 0$. To ensure safe operation, the vector $\mathbf u_i$ should satisfy the following power and energy limits:
\begin{equation}\label{eqn_individual_ori}
	\abovedisplayskip=4pt
	\belowdisplayskip=4pt
	\begin{aligned}
		\mathbf u_i \in \mathcal{U}_i := \left\{ \mathbf u_i \!\in \mathbb{R}^T \!\mid\! \mathbf u_i\! \in [\underline{\mathbf u}_i, \overline{\mathbf u}_i], \ \mathbf L \mathbf u_i\! \in [\underline{\mathbf x}_i, \overline{\mathbf x}_i] \right\}, \forall i \in \mathcal{N}, 
	\end{aligned}
\end{equation}
where $\underline{\mathbf u}_i \!:=\! \big[\underline u_i(t)\big]_{\!t\in \mathcal T}$ and $\overline{\mathbf u}_i \!:=\! \big[\overline u_i(t)\big]_{\!t\in \mathcal T}$ are lower and upper bounds on the charging power, respectively; $\underline{\mathbf x}_i \!:=\! \big[\underline x_i(t)\big]_{\!t\in \mathcal T}$ and $\overline{\mathbf x}_i \!:=\! \big[\overline x_i(t)\big]_{\!t\in \mathcal T}$ represent lower and upper bounds on the cumulative charging energy, respectively; $\mathbf L \in \mathbb{R}^{T \times T}$ is a lower triangular matrix with all its non-zero elements \cg{equal to} the time interval $\Delta t$,
\cg{and $\mathbf L \mathbf u_i$ yields a vector representing the cumulative energy charged to EV $i$ at each time step.}
By defining $\mathbf H := [\mathbf L, -\mathbf L, \mathbf I_T, -\mathbf I_T] \in \mathbb R^{4T\times T}$ and $\mathbf h_i := [\overline{\mathbf x}_i, -\underline{\mathbf x}_i, \overline{\mathbf u}_i, -\underline{\mathbf u}_i] \in \mathbb R^{4T}$, constraint \eqref{eqn_individual_ori} can further be reformulated as follows:
\begin{equation}\label{eqn_individual_ori_H}
	\abovedisplayskip=4pt
	\belowdisplayskip=4pt
	\mathbf u_i \in \mathcal{U}_i = \left\{ \mathbf u_i \in \mathbb{R}^T \mid \mathbf H \mathbf u_i \leq \mathbf h_i \right\},\ \forall i \in \mathcal{N},
\end{equation}
where the set $\mathcal U_i$ defines a polytope that describes the flexibility for adjusting the charging power $\mathbf u_i$, so we call $\mathcal U_i$ the \emph{individual flexibility set} of EV $i$. This set is fully characterized by the parameter $\mathbf h_i$, which is constructed based on the plug-in time, deadline and charging energy demand of EV $i$.~Hence, sharing $\mathbf h_i$ with the aggregator may expose private user preferences, schedules, and energy usage patterns. 

\subsubsection{Aggregate flexibility set}
The \emph{aggregate flexibility set}, which describes the feasible region of the aggregate charging power, can be represented by the Minkowski sum of the individual EV flexibility sets:
\begin{equation}\label{eq:Minksum}
	\abovedisplayskip=4pt
	\belowdisplayskip=4pt
	\mathcal{U} = \oplus_{i \in \mathcal{N}} \mathcal{U}_i = \bigg\{ \mathbf u \in \mathbb{R}^T \mid \mathbf u = \sum_{i \in \mathcal{N}} \mathbf u_i, \ \mathbf u_i \in \mathcal{U}_i \bigg\},
\end{equation}
where $\oplus$ represents the Minkowski sum. 
The computation of this Minkowski sum is NP-hard \cite{nphard}, so directly obtaining the aggregate flexibility set $\mathcal{U}$ is challenging.

 

\vspace{-1mm}
\begin{remark}[Generalization to other DSRs]\label{remark:general_model}
    DSRs that (i) support time-shifting of energy/demand and (ii) admit storage-like linear dynamics for their state, can usually be modeled as the H-polytope form as \eqref{eqn_individual_ori_H}. 
    Examples include flexible loads that can be modeled as \grn{virtual battery systems \cite{7864461}}, such as TCLs, pool pumps, and energy storage systems.
\end{remark}

\subsection{Revisit of Geometric Methods}

Our framework builds upon the principles of geometric aggregation \cite{8063901, 7864461}. The core idea is to inner-approximate the intractable aggregate flexibility set $\mathcal{U}$ with a polytope 
$\widetilde{\mathcal{U}}$ that is computationally efficient to construct. This is achieved by first defining a shared geometric template, the \emph{base set}, which is a convex polytope parameterized by a vector $\mathbf{h}_0$:
\begin{equation}\label{eq:baseSet}
\abovedisplayskip=4pt
	\belowdisplayskip=4pt
	\mathcal{U}_0(\mathbf h_0) = \left\{ \mathbf u \in \mathbb{R}^T \mid \mathbf H \mathbf u \leq \mathbf h_0 \right\}.
\end{equation}
Each participating DSR $i$ then locally computes the largest possible affine transformation of this base set that is contained within its own flexibility set $\mathcal{U}_i$:
\begin{equation}
\abovedisplayskip=4pt
	\belowdisplayskip=4pt
\widetilde{\mathcal{U}}_i(\mathbf{h}_0) =  \bm \gamma_i + \bm \Gamma_i \, \mathcal{U}_0(\mathbf h_0) \subseteq \mathcal{U}_i,
\end{equation}
where $\bm \gamma_i$ and $\bm \Gamma_i$ are the affine transformation parameters. The aggregate flexibility is then approximated by the Minkowski sum of these individual sets, which simplifies to \cite{10490133}:
\begin{equation}
\abovedisplayskip=4pt
	\belowdisplayskip=4pt
	\widetilde{\mathcal{U}}(\mathbf h_0) = \bigg(\sum_{i \in \mathcal{N}}\bm \gamma_i \bigg) + \bigg(\sum_{i \in \mathcal{N}} \bm \Gamma_i \bigg) \, \mathcal{U}_0(\mathbf h_0). \label{eq:agg_flex_set}
\end{equation}
Note the volume of $\widetilde{\mathcal{U}}(\mathbf h_0)$ is a common measure for the characterization performance of aggregate flexibility (the larger the better if the characterized flexibility is realizable). Then, for a given base set, we can formulate the following optimization problem to maximize the characterized aggregate flexibility:
\begin{subequations}\label{eq:opt_org}
 \abovedisplayskip=4pt
 \belowdisplayskip=4pt
    \begin{align}
		\!\!\max_{(\bm \gamma_i, \bm \Gamma_i)_{\forall i \in \mathcal{N}}} \ & \mathrm{vol}\left(\widetilde{\mathcal{U}}(\mathbf h_0)\right) \!=\! \bigg|\mathrm{det}\bigg(\sum_{i\in\mathcal N} \bm \Gamma_i\bigg)\bigg| \!\cdot \!\mathrm{vol}(\mathcal{U}_0(\mathbf h_0)) \\ 
		\mbox{s.t.} \quad\ \, & \bm \gamma_i + \bm \Gamma_i \, \mathcal{U}_0(\mathbf h_0) \subseteq \mathcal{U}_i, \ \ i\in \mathcal N, \label{eq:opt_org:b}
	\end{align}
\end{subequations}
where \eqref{eq:opt_org:b} can be equivalently expressed in linear form using the following lemma \cite{9029363}.

 \vspace{-1mm}
\begin{lemma} \label{lemma_1}
	Suppose there are two polyhedrons \grn{$\mathcal{X} = \{\mathbf x \in \mathbb{R}^{n_x} \mid \mathbf H_x \mathbf x \leq \mathbf h_x\}$ and $\mathcal{Y} = \{\mathbf y \in \mathbb{R}^{n_y} \mid \mathbf H_y \mathbf y \leq \mathbf h_y\}$}, where $\mathbf H_x \in \mathbb{R}^{m_x \times n_x}$, $\mathbf H_y \in \mathbb{R}^{m_y \times n_y}$, and \grn{$\mathcal{X}$} is nonempty. Then, given a vector $\bm \gamma \in \mathbb{R}^{n_y}$ and matrix $\mathbf \Gamma \in \mathbb{R}^{n_y \times n_x}$, it holds that \grn{$\bm \gamma + \mathbf \Gamma \mathcal{X} \subseteq \mathcal{Y}$} if and only if there exists a matrix $\mathbf \Lambda \in \mathbb{R}^{m_y \times m_x}$ such that
	\begin{equation}
		\abovedisplayskip=3pt
		\belowdisplayskip=3pt
		\bm \Lambda \geq \bm 0, \quad \bm \Lambda \mathbf H_x = \mathbf H_y \bm \Gamma,\quad \bm \Lambda  \mathbf h_x \leq \mathbf h_y - \mathbf H_y \bm \gamma. 
	\end{equation} 	
\end{lemma}
\vspace{-1mm}

By substituting Lemma \ref{lemma_1}, problem \eqref{eq:opt_org} becomes
\begin{subequations} \label{eq:opt_lemma}
	\abovedisplayskip=4pt
	\belowdisplayskip=4pt
	\begin{align}
		\max_{(\bm \gamma_i, \bm \Gamma_i)_{\forall i \in \mathcal{N}}} \quad& \ \, \bigg|\mathrm{det}\bigg(\sum_{i\in\mathcal N} \bm \Gamma_i\bigg)\bigg|  \label{eq:opt_lemma_obj}\\
		\mbox{s.t.} \qquad \, & \begin{cases}
			&\!\!\!\!\! \bm \Lambda_i \geq \bm 0, \\
			&\!\!\!\!\! \bm \Lambda_i \mathbf H = \mathbf H \bm \Gamma_i,\\
			&\!\!\!\!\! \bm \Lambda_i  \mathbf h_0 \leq \mathbf h_i - \mathbf H \bm \gamma_i.	
		\end{cases} \ \ \forall i \in \mathcal{N}.
	\end{align}
\end{subequations}
Note that we remove the term $\mathrm{vol}(\mathcal{U}_0(\mathbf h_0))$ from the objective, as it is a constant when $\mathbf h_0$ is given. Obviously, the objective \eqref{eq:opt_lemma_obj} couples different DSRs. As a result, directly solving \eqref{eq:opt_lemma} requires that all DSRs share their sensitive information $\mathbf{h}_i$ with the aggregator, which may cause privacy concerns. 
To address the privacy issue, we linearize the objective using a first-order Taylor expansion about the identity matrix\footnote{The approximation is introduced to enable separability of the objective and enable a federated implementation. Similar linearizations of determinant-based volume metrics are commonly adopted in geometric methods and are empirically validated to provide tight aggregate flexibility representations \cite{10490133}.}, as follows:
\begin{equation}\label{eq:approx}
\abovedisplayskip=4pt
\belowdisplayskip=4pt
\begin{aligned}
\bigg|\mathrm{det}\bigg(\sum_{i\in\mathcal N} \bm \Gamma_i\bigg)\bigg| &\approx \mathrm{Tr}(\sum_{i\in\mathcal N} \bm \Gamma_i) + \mathrm{constant}  \\
	&=  \sum_{i\in\mathcal N} \mathrm{Tr}(\bm \Gamma_i) + \mathrm{constant}.
\end{aligned}
\end{equation}
Then, the optimal solution $(\bm \gamma_i^\star, \bm \Gamma_i^\star)_{\forall i \in \mathcal{N}}$ of problem \eqref{eq:opt_lemma} can be obtained by independently solving the following problem for each individual DSR:
\begin{subequations}\label{eq:opt_EV}
\abovedisplayskip=4pt
\belowdisplayskip=4pt
	\begin{align}
		\max_{\bm \gamma_i, \bm \Gamma_i} \quad & \mathrm{Tr}(\bm \Gamma_i) \\
		\mbox{s.t.} \quad & \bm \Lambda_i \geq \bm 0, \
		\bm \Lambda_i \mathbf H = \mathbf H \bm \Gamma_i, \
		\bm \Lambda_i  \mathbf h_0 \leq \mathbf h_i - \mathbf H \bm \gamma_i.
	\end{align}
\end{subequations}
The aggregate flexibility can be characterized in a decentralized manner: 
each DSR first locally solves its own optimization problem \eqref{eq:opt_EV} using the given base set parameter $\mathbf h_0$, and broadcasts the optimal parameters $(\bm \gamma_i^\star, \bm \Gamma_i^\star)$ to the aggregator. The aggregate then formulates a set $\widetilde{\mathcal{U}}^\star(\mathbf h_0) := \left(\sum_{i \in \mathcal{N}}\bm \gamma_i^\star\right) + \left(\sum_{i \in \mathcal{N}} \bm \Gamma_i^\star \right) \, \mathcal{U}_0(\mathbf h_0)$ as a high-quality inner approximation of the aggregate flexibility set $\mathcal{U}$.

The efficacy of this approach is critically dependent on the choice of the base set. In existing static implementations, the parameter $\mathbf{h}_0$ is treated as a fixed, pre-determined input, often chosen by heuristics such as the average of all $\mathbf{h}_i$ \cite{10490133}. This static-template design is the source of the performance bottleneck. Thus, a fundamental challenge is how to overcome the static limitation by optimizing the parameter $\mathbf{h}_0$ itself.

\subsection{Base Set Optimization}

A straightforward solution to the previous challenge is to treat the base set parameter $\mathbf{h}_0$ as an optimizable decision variable, resulting in the following optimization problem:
\begin{subequations} \label{eqn_centralized}
\abovedisplayskip=4pt
	\belowdisplayskip=4pt
\begin{align} 
	\max_{\mathbf h_0} \quad &  \bigg| \mathrm{det} \bigg(\sum_{i\in\mathcal N} \boldsymbol{\Gamma}_i^\star\bigg) \bigg| \cdot \mathrm{vol}(\mathcal{U}_0(\mathbf h_0)),  \\
	\mbox{s.t.} \quad & \mathrm{int}(\mathcal{U}_0(\mathbf h_0)) \neq \varnothing \label{eqn_constraint_noempty}, \\
	& (\bm \gamma_i^\star, \bm \Gamma_i^\star)~\text{is a solution of \eqref{eq:opt_EV}}, \ \forall i \in \mathcal{N}. \label{eqn_argmax}
\end{align}    
\end{subequations}
The objective maximizes the volume of the inner approximation of the aggregate flexibility set $\widetilde{\mathcal{U}}(\mathbf h_0)$ defined in \eqref{eq:agg_flex_set}. 
\cg{Constraint \eqref{eqn_constraint_noempty} enforces that the interior of $\mathcal{U}_0$ is non-empty.}
Constraint \eqref{eqn_argmax} defines the lower-level problems whose optimal solutions $\bm \Gamma_i^\star$ (and  $\bm \gamma_i^\star$) for all $i\in\mathcal N$ are functions of the base set parameter $\mathbf h_0$.


Solving problem \eqref{eqn_centralized} is challenging, because i) it is a bilevel problem due to constraint \eqref{eqn_argmax}, ii) the objective function is non-concave because both $\bm \Gamma_i^\star$ and $\mathbf h_0$ are variables, iii) accurately computing the volume of the base set $\mathcal{U}_0(\mathbf h_0)$ is computationally demanding, and iv) the constraint cannot be directly handled by existing optimization solvers. Together, these challenges render problem \eqref{eqn_centralized} intractable for off-the-shelf solvers.
Moreover, this centralized formulation requires DSRs to share the vectors $\mathbf h_i$ for all $i\in\mathcal N$ with the~aggregator. Since $\mathbf h_i$ encodes user preferences, schedules, and energy usage patterns, such disclosure directly exposes sensitive operational information and undermines user privacy.

\section{Federated Base Set Optimization}\label{sec:method}
To address the above challenges, we propose a federated method to optimize the base set. First, we reformulate the base set optimization problem \eqref{eqn_centralized} into an unconstrained learning task. Second, we present a decentralized gradient-based algorithm tailored for this new formulation by exploiting the gradient's decomposable structure. Finally, we develop the differentiable modules required to compute the gradient.

\subsection{Unconstrained Reformulation}
We first provide an unconstrained reformulation for the base set optimization problem \eqref{eqn_centralized}. This reformulation employs the concept of \emph{optimal solution functions} to transform the problem into an unconstrained form.
Specifically, constraint \eqref{eqn_constraint_noempty} enforces that the optimized $\mathbf{h}_0$ corresponds to a base set $\mathcal{U}_0$ with non-empty interior. This constraint can be safely approximated by the following solution function over an L2-norm projection, as follows:
\begin{subequations} \label{eqn_projection}
	\abovedisplayskip=4pt
	\belowdisplayskip=4pt
 \begin{align}
\widetilde{\mathbf h}_0(\mathbf h_0) = \argmin_{\mathbf h_0', \mathbf u}\quad & \Vert \mathbf h_0 -\mathbf h_0' \Vert_2^2,  \\
\text{s.t.} \quad& \mathbf H \mathbf u + \epsilon \bm 1 \leq \mathbf h_0', \label{eqn_proj_constraint}
\end{align}
 \end{subequations}
where $\epsilon>0$ is a small positive constant, ensuring that the interior $\mathcal{U}_0\big(\widetilde{\mathbf h}_0\big)$ is non-empty. The solution $\widetilde{\mathbf h}_0(\mathbf h_0)$ must be feasible for constraint \eqref{eqn_constraint_noempty}. Similarly, constraint \eqref{eqn_argmax} is defined as an optimal solution of problem \eqref{eq:opt_EV}, so it can be directly replaced by the corresponding solution function $\boldsymbol{\Gamma}_i^\star\big(\widetilde{\mathbf h}_0\big)$. By substituting the previous solution functions, we can formulate an unconstrained problem to approximate \eqref{eqn_centralized}:
\begin{equation} \label{eqn_centralized_2}
\abovedisplayskip=4pt
\belowdisplayskip=4pt
\max_{\mathbf h_0} ~ J (\mathbf{h}_0) =  \ell (\mathbf{h}_0) \cdot \phi(\mathbf{h}_0),
\end{equation}
where functions $\ell (\mathbf{h}_0)$ and $\phi(\mathbf{h}_0)$ are defined as
\begin{equation} 
 \begin{aligned}
    & \ell (\mathbf{h}_0) = \bigg|\mathrm{det} \bigg(\sum_{i\in\mathcal N} \boldsymbol{\Gamma}_i^\star\left(\widetilde{\mathbf h}_0(\mathbf h_0)\right)\bigg)\bigg|, \label{eqn_ell} \\
    & \phi(\mathbf{h}_0) = \mathrm{vol}\left(\mathcal{U}_0\left(\widetilde{\mathbf h}_0(\mathbf h_0)\right)\right).  
 \end{aligned}
\end{equation}
This unconstrained problem can be efficiently solved  by gradient-based algorithms. Note that a small perturbation, $\epsilon \bm 1$, is introduced into constraint \eqref{eqn_proj_constraint} to ensure the projected solution, $\widetilde{\mathbf h}_0$, remains feasible for the original optimization problem \eqref{eqn_centralized}. Since this modification only marginally contracts the feasible region, any resulting optimality gap is expected to be minimal.

\subsection{Federated Gradient-based Algorithm} \label{sec:decentralized_gradient}
While the unconstrained formulation \eqref{eqn_centralized_2} can be solved using gradient-based methods, a naive implementation would require a central aggregator to have access to all DSRs' private information ($\mathbf{h}_i$) to compute the gradient of $J(\mathbf{h}_0)$. To enable a truly federated solution, we design a federated algorithm based on the decomposable structure of this gradient.

\begin{figure}[ht]
	\centering	{\includegraphics[width=0.9\columnwidth]{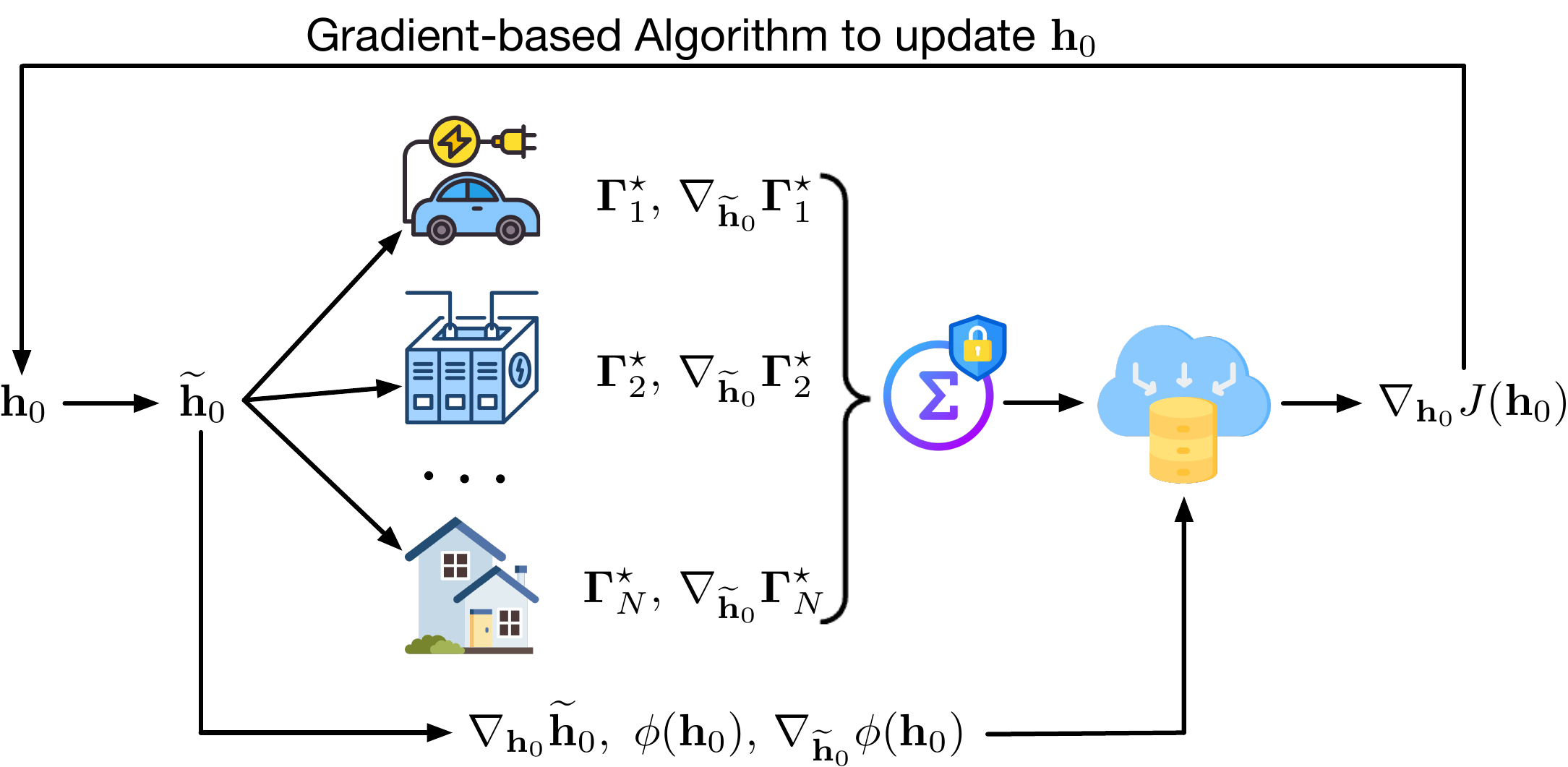}}
	\vspace{-3mm}
	\caption{The federated gradient-based algorithm for base set optimization.}
	 \vspace{-3mm}
	\label{fig_summary}
\end{figure}

The gradient of the objective function $J(\mathbf{h}_0)$ with respect to the decision variable $\mathbf{h}_0$, $\nabla_{\mathbf h_0}J (\mathbf{h}_0)$, can be computed using the chain rule:
\begin{equation}\label{eq:dJ_dh0}
\abovedisplayskip=3pt
\belowdisplayskip=3pt
	\nabla_{\mathbf h_0}J (\mathbf{h}_0) = \nabla_{\widetilde{\mathbf h}_0}J (\widetilde{\mathbf h}_0) \cdot \nabla_{{\mathbf h}_0}\widetilde{\mathbf h}_0({\mathbf h}_0)
\end{equation}
The first term on the right-hand side further expands as:
\begin{equation}\label{eq:dJ_dhtilde}
\abovedisplayskip=3pt
\belowdisplayskip=3pt
	\nabla_{\widetilde{\mathbf h}_0}J (\widetilde{\mathbf h}_0) = \ell(\widetilde{\mathbf h}_0) \cdot \nabla_{\widetilde{\mathbf h}_0}\phi(\widetilde{\mathbf h}_0) + \phi(\widetilde{\mathbf h}_0) \cdot \nabla_{\widetilde{\mathbf h}_0} \ell(\widetilde{\mathbf h}_0).
\end{equation}
The key insight lies in the decomposable structure of the terms $\ell(\widetilde{\mathbf h}_0)$ and $\nabla_{\widetilde{\mathbf h}_0} \ell(\widetilde{\mathbf h}_0)$, both of which involve summations over all DSRs. Specifically, the gradient can be expressed as:
\begin{equation} \label{eq:grad_ell_decomposed}
\abovedisplayskip=3pt
\belowdisplayskip=3pt
	\begin{aligned}
		\nabla_{\widetilde{\mathbf h}_0} \ell(\widetilde{\mathbf h}_0) = \ell(\widetilde{\mathbf h}_0) \cdot \mathrm{Tr}\bigg[\bigg(\sum_{i\in \mathcal N} \boldsymbol{\Gamma}_i^\star\bigg)^{-1} \bigg(\sum_{i\in \mathcal N} \nabla_{\widetilde{\mathbf h}_0} \boldsymbol{\Gamma}_i^\star\bigg)\bigg].
	\end{aligned}
\end{equation}
Then, the overall gradient $\nabla_{\mathbf h_0}J (\mathbf{h}_0)$ can be computed in a federated manner, as shown in Fig. \ref{fig_summary}: 
\begin{itemize}
	\item \textbf{Local Computation:} Given the shared $\widetilde{\mathbf h}_0$, DSR $i$ can independently compute its optimal transformation $\boldsymbol{\Gamma}_i^\star(\widetilde{\mathbf h}_0)$ and its local gradient $\nabla_{\widetilde{\mathbf h}_0} \boldsymbol{\Gamma}_i^\star(\widetilde{\mathbf h}_0)$.
	\item \textbf{Central Computation:} The aggregator computes all the other terms that do not depend on private DSR data, including the volume $\phi(\widetilde{\mathbf h}_0)$, its gradient $\nabla_{\widetilde{\mathbf h}_0}\phi(\widetilde{\mathbf h}_0)$, and the projection gradient $\nabla_{{\mathbf h}_0}\widetilde{\mathbf h}_0({\mathbf h}_0)$. Upon receiving $(\boldsymbol{\Gamma}_i^\star, \nabla_{\widetilde{\mathbf h}_0} \boldsymbol{\Gamma}_i^\star)$ from all DSRs, the aggregator computes $\ell(\widetilde{\mathbf h}_0)$ and $\nabla_{\widetilde{\mathbf h}_0} \ell(\widetilde{\mathbf h}_0)$. Crucially, since these terms involve only aggregated sums, they can be computed using secure summation protocols that preserve anonymity and substantially limit the disclosure of user contributions. Finally, the overall gradient can be computed according to \eqref{eq:dJ_dh0}-\eqref{eq:grad_ell_decomposed}.
\end{itemize}
This process allows for the collaborative optimization of the base set without access to private parameters $\mathbf{h}_i$ for all $i\in \mathcal N$.

\subsection{Computation of the Gradient's Components} \label{sec:gradient_computation}

While the gradient's decomposable structure provides the theoretical foundation for our federated algorithm, its practical implementation relies on the computation of its constituent components. This section details the methodologies we employ to compute these components.

\subsubsection{Gradient $\nabla_{\mathbf h_0}\widetilde{\mathbf h}_0({\mathbf h}_0)$}
This term is hard to explicitly formulate because it involves an optimization problem \eqref{eqn_projection}. Nevertheless, since this problem is convex and satisfies regularity conditions, gradient $\nabla_{{\mathbf h}_0}\widetilde{\mathbf h}_0({\mathbf h}_0)$ can be obtained by applying the Karush-Kuhn-Tucker (KKT) conditions and the implicit function theorem \cite{ab4575cf4ab642d89d2a88041b41675e}. Appendix~\ref{app:projection} provides the analytical formulation of this gradient.

\subsubsection{Function $\ell(\widetilde{\mathbf h}_0)$ and gradient $\nabla_{\widetilde{\mathbf h}_0} \ell(\mathbf h_0)$}
The former is directly computed according to \eqref{eqn_ell}. The latter is calculated based on \eqref{eq:grad_ell_decomposed}, which relies on the gradient $\nabla_{\widetilde{\mathbf h}_0} \boldsymbol{\Gamma}_i^\star$. 
Since this gradient also involves a convex optimization problem \eqref{eq:opt_EV}, it can also be computed based on KKT conditions and the implicit function theorem. \cg{Appendix~\ref{app:setContain} provides the analytical formulation of this gradient}.

\subsubsection{Base set volume $\phi(\mathbf h_0)$ and gradient $\nabla_{\widetilde{\mathbf h}_0}\phi(\mathbf h_0)$}

Calculating the volume of a high-dimensional polytope, such as $\phi(\mathbf h_0)$, is generally intractable \cite{doi:10.1137/0217060}. Existing volume approximation methods typically rely on Monte Carlo sampling \cite{BAROT201755, 10.1145/3584182}, where the volume is inferred based on the proportion of sampled points within the polytope. However, these methods are not suitable for computing gradients, as they do not provide a differentiable formulation of the volume. Moreover, maintaining accurate estimates typically requires a prohibitively large number of samples, resulting in significant computational overhead. To address these limitations, we derive analytical expressions for both the volume and its gradient, which admit a predictable polynomial computational complexity (see details in Section~\ref{sec:complexity}). For an arbitrary base set parameter $\mathbf h_0$, the volume of the base set defined in \eqref{eq:baseSet} can be expressed as 
\begin{equation}\label{eq:vol_def}
\abovedisplayskip=4pt
\belowdisplayskip=4pt
    \mathrm{vol}(\mathcal U_0(\mathbf h_0)) = \int_{\mathbf u\in\mathbb R^T} \mathbb{I}\big\{\mathbf H\mathbf u \le \mathbf h_0\big\} \, \mathrm{d}\mathbf u,
\end{equation}where $\mathbb{I}\{\mathfrak C\}$ is the indicator function (equal to 1 when the condition $\mathfrak C$ holds and 0 otherwise). This $T$-dimensional integral admits a nested (and iterated) one-dimensional representation. Differentiating that representation with~respect to the parameter $\mathbf h_0$ yields the gradient $\nabla_{\mathbf h_0}\mathrm{vol}(\mathcal U(\mathbf h_0))$. \grn{Further details~on the nested forms are provided in Appendix \ref{app:vol}.} Both the volume and its gradient can then be estimated efficiently via a recursive numerical  integration scheme, as discussed below.

\begin{remark}[Volume and gradient estimation]\label{remark:volume}
\jq{Efficient estimation of the volume of a high-dimensional polytope and its gradient with respect to parameters $\mathbf h_0$, both as functions of $\mathbf h_0$, is challenging. Instead of naively applying  methods that directly perform numerical integration/Monte Carlo simulation in $\mathbb R^T$, in Appendix \ref{app:vol}, we devise a nested representation for the volume and its gradient, that only relies on \grn{a nested sequence of one-dimensional integrals over state-dependent intervals, enabling a dimension-by-dimension computation.~Building on this structure, Appendix~\ref{app:mc_integration} presents a recursive numerical scheme that estimates both the volume and its gradient by~discretizing each integral over a fixed and bounded interval, and reusing intermediate quantities, yielding efficient computation.}}
\end{remark}

\subsection{Computational Complexity and Communication Cost}\label{sec:complexity}
We now characterize the per-iteration computational cost of the proposed algorithm. In each iteration, the computational workload is divided between the aggregator and all DSRs:

\begin{enumerate}	
	\item \textbf{Aggregator side.} The aggregator performs the projection step \eqref{eqn_projection}, and estimates the base-set volume and its gradient. The projection problem is a convex problem with $O(T)$ variables and constraints, solvable in polynomial time using a primal–dual interior-point method. Following Appendices~\ref{app:vol} and~\ref{app:mc_integration}, the $T$-dimensional volume integral \eqref{eq:vol_def} and its gradient are computed through a recursive sequence of one-dimensional Riemann sums. Specifically, at each dimension $t$, the state space is discretized into $K_t$ grid points, and each grid point aggregates over $K_{t-1}$ points from the previous step, resulting in a total runtime of $O(TK_{\max}^2)$ with $K_{\max}=\max_{t\in\mathcal T} K_t$. Hence, the overall per-iteration complexity on the aggregator side is $O(TK_{\max}^2)+\mathrm{poly}(T)$.
	
	\item \textbf{DSR side.} Each DSR solves its local set-containment problem \eqref{eq:opt_EV}, which is a linear program with $O(T^2)$ variables and constraints and can be solved in polynomial time with an interior-point method. Since all DSRs operate in parallel, the per-iteration wall-time remains roughly constant with $N$, with per-agent complexity $\mathrm{poly}(T^2)$.
\end{enumerate}

We then analyze the per-iteration communication cost of the proposed algorithm. In each iteration, the aggregator broadcasts the current base-set parameter $\mathbf h_0$ to all DSRs, incurring a downlink cost of $O(T)$ scalars that is independent of $N$. Each DSR then returns the optimal affine transformation $\bm\Gamma_i$ and the associated gradient information. The uplink communication cost per DSR scales as $O(\mathrm{dim}(\bm\Gamma_i))$, which is $O(T^2)$ in the worst case, and the total uplink cost therefore scales linearly with $N$. As only aggregated quantities are required and all local computations are parallelized, the overall communication overhead remains lightweight and scalable, while preserving user privacy by design.

\subsection{Applicability for Heterogeneous DSRs}
While the above pipeline is formulated for resources whose flexibility sets share a common structure, where variability arises primarily from the right-hand-side parameters $\mathbf h_i$ and all left-hand-side matrices $\mathbf H$ are identical, the proposed approach is inherently general and can accommodate a broader range of DSRs. In particular, for devices such as TCLs whose~feasible regions depend on both the left-hand and right-hand parameters \cite{7864461}, the same optimization pipeline remains applicable.

Specifically, similar to \eqref{eqn_individual_ori_H}, the flexibility set of TCL $i$ has a left-hand matrix, denoted as $\mathbf H_i^\mathrm{TCL}$, which is fully~determined by the dissipation rate $a_i$. It can be shown, using Neumann series \cite{6572301}, that $\mathbf H_i^\mathrm{TCL}$ is polynomial in $a_i$. The corresponding right-hand-side vector, denoted by $\mathbf h_i^\mathrm{TCL}$, incorporates the modified state and input limits consistent with the parameter $a_i$. Thus, an \emph{augmented} base-set parameter $[a_0, \mathbf h_0^\mathrm{TCL}]$ can be treated as the variable to be optimized. The objective function and its gradient retain structures similar to those in \eqref{eqn_centralized_2} and \eqref{eq:dJ_dh0}, with corresponding gradient components described in Section~\ref{sec:gradient_computation}. These quantities can be derived analogously, allowing the optimization to be performed within the same federated base-set framework without modification.

Finally, the proposed framework naturally extends to heterogeneous populations of DSRs, including EVs, TCLs, and other flexible loads, as it does not require identical constraint matrices across participants. Each DSR is modeled by its own polyhedral flexibility set with potentially distinct left-hand and right-hand parameters. Such heterogeneity is handled locally through the set-containment problems, while the aggregator optimizes a shared base set to approximate the collection of individual flexibility sets via affine transformations. When heterogeneity is sufficiently pronounced that a single base set becomes overly restrictive, the framework admits a multi–base-set extension, in which DSRs are grouped by similarity and each group is assigned its own optimized base set. Aggregation then proceeds independently within each group, and the resulting group-level flexibility sets can be either directly used by the power system operator or combined at a higher level, while preserving the same optimization pipeline and privacy properties.

\vspace{-1mm}
\begin{remark}[Applicability for non-linear and discrete dynamics]
	The proposed framework can be extended beyond settings where individual DSR models are linear or continuous at the device level. For DSRs with nonlinear dynamics, convex~polyhedral inner approximations of the resulting nonconvex but compact and continuous sets have been established in the literature \cite{HiriartUrruty1996, BallMilman1999}, and can be directly used within the framework.  For DSRs with discrete control actions (e.g., ON/OFF switching in TCLs), continuous convex flexibility sets can be obtained by modeling flexibility at an appropriate temporal or population scale, where binary actions admit meaningful continuous representations (e.g., duty cycles or population-level averages) \cite{7864461, 10057479}. As long as a DSR’s flexibility can be represented by a convex polytope in H-representation, our framework remains applicable.
	
\end{remark}

\section{Federated Aggregation and Disaggregation} \label{sec:protocols}


\grn{In this section, we utilize the \emph{block-separable} structure of the aggregation/disaggregation methods in \cite{10490133} to design a decentralized implementation in a federated setting. Together with the proposed federated base set optimization methods, we then obtain an end-to-end federated framework for demand flexibility aggregation.
}

\subsection{Federated Aggregation} \label{sec:agg_protocol}

The goal of the aggregation protocol is to compute the aggregate flexibility set $\widetilde{\mathcal{U}}(\mathbf{h}_0)$ without the aggregator accessing individual DSRs' private data $\mathbf{h}_i$. The detailed steps are:

\begin{enumerate}
	\item \textbf{Step 1: Broadcast by Aggregator.} The aggregator broadcasts the optimized base set to all participating DSRs.
	
	\item \textbf{Step 2: Local Computation at DSRs.} Upon receiving the base set, each DSR $i$ locally solves problem \eqref{eq:opt_EV} to find its optimal affine transformation parameters $(\bm{\gamma}_i^\star, \bm{\Gamma}_i^\star)$. This problem is convex and can be solved efficiently by each DSR using its own information $\mathbf{h}_i$.
	
	\item \textbf{Step 3: Central Aggregation.} The aggregator collects the optimal parameters from all DSRs and computes the aggregate transformation parameters by:
	\begin{equation}
    \abovedisplayskip=3pt
	\belowdisplayskip=3pt
		\mathbf{s}^\text{agg} = \sum_{i \in \mathcal{N}} \bm{\gamma}_i^\star, \quad \mathbf{S}^\text{agg} = \sum_{i \in \mathcal{N}} \bm{\Gamma}_i^\star.
	\end{equation}
	This can also be implemented via black-box (secure) summation \cite{10490133}, so individual identities and per-DSR parameters are not revealed. The final aggregate flexibility set is then constructed as:
	\begin{equation}
    \abovedisplayskip=3pt
	\belowdisplayskip=1pt
		\widetilde{\mathcal{U}}(\mathbf{h}_0) = \mathbf{s}^\text{agg} + \mathbf{S}^\text{agg}\, \mathcal{U}_0(\mathbf{h}_0).
	\end{equation}
\end{enumerate}

\subsection{Federated Disaggregation} \label{sec:disagg_protocol}

Once the aggregate flexibility set $\widetilde{\mathcal{U}}(\mathbf{h}_0)$ is constructed and used for grid services, the aggregator may receive a target aggregate power profile, $\mathbf{u}^\text{agg} \in \widetilde{\mathcal{U}}(\mathbf{h}_0)$, that needs to be disaggregated back to individual DSRs. The disaggregation protocol must ensure that each DSR receives a feasible local profile, again without compromising privacy. The detailed steps for disaggregation are as follows: 
\begin{enumerate}
	\item \textbf{Step 1: Central Computation by Aggregator.} The aggregator first computes a \emph{base profile} $\mathbf{u}_0$. This is done by reversing the aggregate transformation:\footnote{While a formal proof of the non-singularity of $\mathbf{S}^\text{agg}$ is not provided, we empirically observe that this matrix is consistently invertible, provided that at least one DSR is connected in each time slot. This is often the case when aggregating a large number of DSRs.}
	\begin{equation}
    \abovedisplayskip=3pt
\belowdisplayskip=3pt
		\mathbf{u}_0 = (\mathbf{S}^\text{agg})^{-1} (\mathbf{u}^\text{agg} - \mathbf{s}^\text{agg}).
	\end{equation}
Then, the aggregator broadcasts this base profile $\mathbf{u}_0$ to all participating DSRs.

	\item \textbf{Step 2: Local Profile Reconstruction at DSRs.} Upon receiving $\mathbf{u}_0$, each DSR $i$ uses its own locally stored optimal transformation parameters $(\bm{\gamma}_i^\star, \bm{\Gamma}_i^\star)$ to reconstruct its individual dispatch profile $\mathbf{u}_i$:
	\begin{equation}
    \abovedisplayskip=4pt
\belowdisplayskip=4pt
		\mathbf{u}_i = \bm{\gamma}_i^\star + \bm{\Gamma}_i^\star \mathbf{u}_0.
	\end{equation}
	This reconstructed profile $\mathbf{u}_i$ is guaranteed to be feasible, i.e., $\mathbf{u}_i \in \mathcal{U}_i$, and the sum of all individual profiles will equal the target aggregate profile: $\sum_i \mathbf{u}_i = \mathbf{u}^\text{agg}$.
\end{enumerate}

\section{Case Study} \label{sec:simulation}
In this section, we will characterize the aggregate flexibility of EVs as an example to demonstrate the efficacy of our proposed federated framework. 

\vspace{-1mm}
\subsection{Simulation Setup}
\subsubsection{Parameter settings}
We consider a finite time horizon with each time interval being 1 hour ($\Delta t = 1$), and an EV fleet consisting of 50 vehicles ($N=50$). The parameter intervals of EVs are summarized in Table~\ref{tab:parameter} (the parameters of a specific EV are uniformly sampled from these intervals). The energy demand for EV $i$ is computed by first determining its maximum feasible charge, given by $u_i^\mathrm{max}(t_i^\mathrm{d} - t_i^\mathrm{p})$, and then scaling it by a random number drawn uniformly from $[0.2, 0.5]$. The parameter $\epsilon$ in constraint \eqref{eqn_proj_constraint} is set to $1\times10^{-5}$, which is sufficiently small to ensure strict feasibility without affecting the geometry of the base set.
\begin{table}[!htbp]
\renewcommand{\arraystretch}{1.3}
\vskip-3mm
\footnotesize
\caption{EV Parameters}
\vspace{-14pt}
\begin{center}
    \begin{tabular}{ ccc } 
        \hline
        Parameter& Description & Value \\ 
        \hline
        $x_i^\mathrm{max}$ & Battery capacity of EV $i$  & $[40, 60]$ kWh \\
        $u_i^\mathrm{max}$ & Maximum charging rate of EV $i$  & $[6, 9]$ kW \\
        $u_i^\mathrm{min}$ & Minimum charging rate of EV $i$  & $[-9, -6]$ kW \\
        $t_i^\mathrm{p}$ & Plug-in time of EV $i$  & $[1, T\!-\!1]$ \\
        $t_i^\mathrm{d}$ & Deadline of EV $i$  & $[t_i^\mathrm{p}\!+\!1, T]$ \\
        $x_i^\mathrm{init}$ & Initial state-of-charge of EV $i$  & $[0, 0.4x_i^\mathrm{max}]$ kWh \\
        \hline
    \end{tabular}
    \label{tab:parameter}
\end{center}
\vspace{-3mm}
\end{table}

\subsubsection{Initialization and step-size selection}
The gradient-based algorithm in Section~\ref{sec:method} is initialized with an all-one vector, providing a feasible starting point\footnote{This initialization guarantees feasibility of~\eqref{eqn_constraint_noempty}, since the origin $\mathbf u=\mathbf 0$ strictly satisfies $\mathbf H\mathbf u \le \mathbf 1$, ensuring that the base set has a nonempty interior.}. The step-size selection follows a standard \emph{backtracking line search method} \cite{ab4575cf4ab642d89d2a88041b41675e} to ensure monotonic ascent of the objective function. 



\subsubsection{Benchmarks}
We implement the following two methods for comparison:
\begin{itemize}
    \item \textsf{AVG}: Geometric method used in \cite{10490133}. The base set is predetermined with its parameter defined as the average of all EVs' parameters, i.e., $\mathbf h_0^\mathrm{avg} := \sum_{i\in\mathcal N}\mathbf h_i$.
    \item \textsf{Proposed}: The proposed method in Section \ref{sec:method}, where an optimized base set parameter $\mathbf h_0^\mathrm{opt}$ is applied.
\end{itemize}
In addition to the baseline \textsf{AVG}, we also include a centralized full-information benchmark in Section~\ref{sec:practical_cases}, where the aggregator directly solves the original optimization problems with access to all individual flexibility sets. This serves as a best-case performance reference rather than a deployable baseline.


\subsection{Volume of Aggregate Flexibility Set}
As a larger volume of the aggregate flexibility set signifies greater flexibility potential, we introduce a \emph{normalized volume ratio}, $r(\mathbf h_0^\mathrm{opt}, \mathbf h_0^\mathrm{avg})$, to quantify the performance improvement. This metric captures the geometric mean of the scaling factor per dimension and is defined as:
\begin{equation}
\abovedisplayskip=1pt
\belowdisplayskip=2pt
    r(\mathbf h_0^\mathrm{opt}, \mathbf h_0^\mathrm{avg}) = \left(\frac{\mathrm{vol}\big(\widetilde{\mathcal{U}}(\mathbf h_0^\mathrm{opt})\big)}{\mathrm{vol}\big(\widetilde{\mathcal{U}}(\mathbf h_0^\mathrm{avg})\big)}\right)^{\!\! 1/T},
\end{equation}
where $\mathrm{vol}\big(\widetilde{\mathcal{U}}(\mathbf h_0^\mathrm{opt})\big)$ and $\mathrm{vol}\big(\widetilde{\mathcal{U}}(\mathbf h_0^\mathrm{avg})\big)$ represent the volumes of the aggregate flexibility sets under base sets $\mathcal{U}_0(\mathbf h_0^\mathrm{opt})$ and $\mathcal{U}_0(\mathbf h_0^\mathrm{avg})$, respectively. Obviously, $r(\mathbf h_0^\mathrm{opt}, \mathbf h_0^\mathrm{avg})>1$ means that the proposed method characterizes a larger aggregate flexibility set.

\begin{figure}[ht]
     \vspace{-2mm}
    \centering	{\includegraphics[width=0.9\columnwidth]{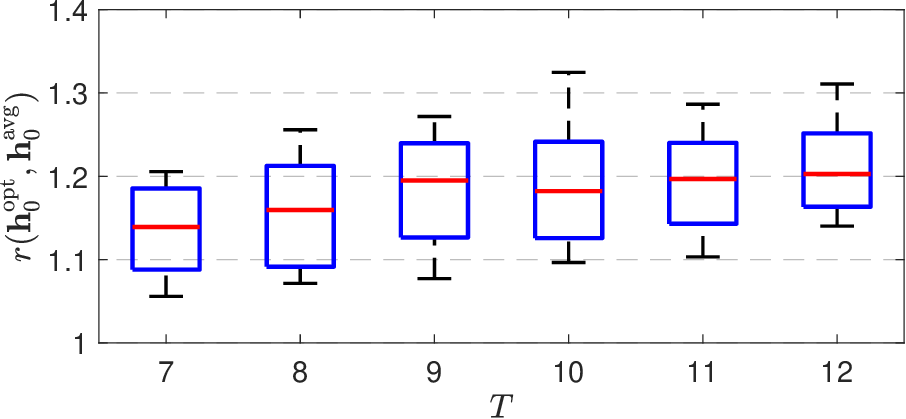}}
    \vspace{-3mm}
    \caption{Distributions of normalized volume ratios under different time-horizon lengths. The whiskers delimit the interdecile range, the boxes delimit the interquartile range, and the red lines represent the medians of each distribution.}
     \vspace{-2mm}
    \label{volumeRatio}
\end{figure}

Fig. \ref{volumeRatio} illustrates the distribution of $r(\mathbf h_0^\mathrm{opt}, \mathbf h_0^\mathrm{avg})$ under different time-horizon lengths. For each length $T$, we run 50 trials with randomly generated EV populations. Across all horizons and trials, the value of $r(\mathbf h_0^\mathrm{opt}, \mathbf h_0^\mathrm{avg})$ is always larger than 1, confirming that the optimized base set always yields a larger inner approximation than the average-based baseline. Median ratios lie around 1.14–1.20, meaning the proposed method enlarges the aggregate set by roughly 14–20\% per dimension on average. The medians (and whiskers) generally increase with $T$, indicating that the benefit of optimizing the base set grows with the time-horizon dimensionality. Overall, base-set optimization unlocks substantially more flexibility than using a pre-specified base set, with gains that become more pronounced for longer horizons.

\begin{figure}[ht]
	\vspace{-2mm}
	\centering	{\includegraphics[width=0.9\columnwidth]{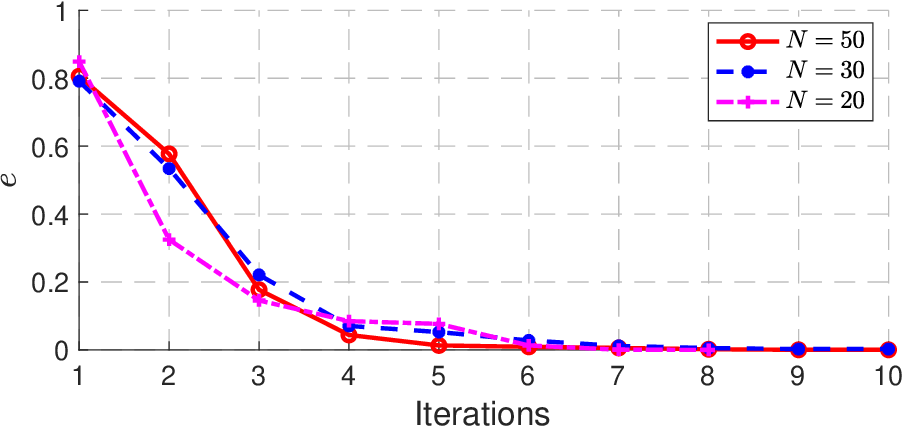}}
	\vspace{-3mm}
	\caption{Convergence trajectories of the normalized error $e$ for different numbers of EVs $N=20, 30, 50$.}
	\vspace{-2mm}
	\label{convergence}
\end{figure}

To further evaluate the efficiency of the proposed federated base-set optimization framework, we investigate its convergence behavior and scalability with respect to the number of participating agents. The convergence performance is quantified using the \emph{normalized suboptimality} defined as
\begin{equation}
	\abovedisplayskip=3pt
	\belowdisplayskip=3pt
	e = \frac{J(\mathbf h_0^{\mathrm{opt}}) - J\big(\mathbf h_0^{(k)}\big)}{J(\mathbf h_0^{\mathrm{opt}})},
\end{equation}where $J\big(\mathbf h_0^{(k)}\big)$ denotes the objective value at iteration $k$. This metric measures the relative distance from the current iterate to the optimal solution. Fig.~\ref{convergence} illustrates the convergence trajectories of $e$ for different numbers of EVs~($N = 20, 30, 50$). Only the first ten iterations are displayed, as the objective values in all cases stabilize within this range, indicating convergence or near-convergence. The average per-iteration computation time at the aggregator side is 42.7~s, 37.4~s, and 44.1~s for $N=20$, $30$, and $50$, respectively, while the corresponding~average per-iteration computation time at the DSR side is 9.8~s, 11.2~s, and 7.6~s. These results indicate that both the convergence speed and per-iteration computational cost exhibit limited sensitivity to different values of $N$, confirming that the proposed algorithm scales efficiently with the number of participating EVs.

\subsection{Integrated Tests with Flexibility Use Cases}\label{sec:practical_cases}

We further evaluate the practical benefits of our method by applying the characterized aggregate flexibility sets to two downstream optimization tasks: electricity cost minimization and peak power minimization. By comparing the optimal solutions achieved with the flexibility sets characterized by different methods, we can assess how the base set optimization enhances EV flexibility to deliver cost-effective grid services.

\subsubsection{Peak power minimization} 
This task considers a commercial building that shares a single service meter with a co-located EV charging facility. The objective is to exploit the connected EVs’ aggregate flexibility to minimize the metered peak demand (e.g., to reduce demand charges).
It can be formulated as the following optimization problem:
\begin{equation}\label{eq:pp_min}
 \abovedisplayskip=3pt
 \belowdisplayskip=3pt
    \min_{\mathbf u} \ ||\mathbf u + \mathbf p||_\infty \ \ \mathrm{s.t. } \   \mathbf u\in \widetilde{\mathcal U},
\end{equation}
where $\mathbf p\in \mathbb R^T$ (kW) is the original load profile of a given commercial building; 
$\mathbf u$ serves as the flexible load that can be adjusted within the aggregate flexibility set $\widetilde{\mathcal U}$. The objective is to minimize the maximum power across the time horizon. The constraint ensures the power adjustment is realizable. 
 
We use historical power consumption data from a commercial building located in San Francisco, CA, obtained from the ComStock dataset~\cite{parker2023comstock}, as the load profile (364 instances throughout the year 2018). The following \emph{centralized optimization} problem is also solved as a best-case benchmark: 
\begin{equation}\label{eq:pp_min_centralized}
\abovedisplayskip=3pt
\belowdisplayskip=3pt
    \bar P^\star = \min_{\mathbf u_i, \forall i \in \mathcal{N}} \ \bigg|\bigg| \sum_{i\in \mathcal N}\mathbf u_i + \mathbf p \bigg|\bigg|_\infty, \ \ \text{s.t. }\  \mathbf u_i \in \mathcal U_i, \, \forall i \in \mathcal N,
\end{equation}
where the flexibility set for each individual EV, $\mathcal U_i$, is exactly known to the aggregator.



\begin{figure}[ht]
	\centering	{\includegraphics[width=0.9\columnwidth]{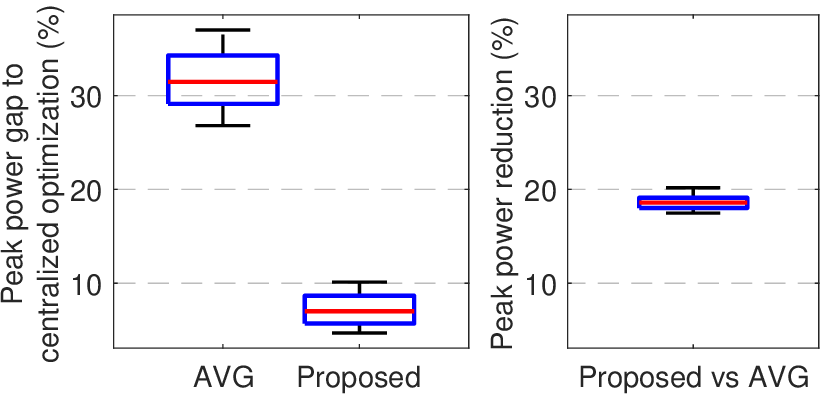}}
	\vspace{-3mm}
	\caption{Distributions of the peak power gaps and improvements for the peak power minimization problem~\eqref{eq:pp_min} under different methods.}
	 \vspace{-4mm}
	\label{gapDistributions_1}
\end{figure}

Fig.~\ref{gapDistributions_1} illustrates i) the relative peak power gaps of the \textsf{AVG} and \textsf{Proposed} methods against the best-case benchmark, computed as $\big(\bar P(\mathbf{h}_0^{\mathrm{avg}}) - \bar P^\star\big)/{\bar P^\star}$ and $\big(\bar P(\mathbf{h}_0^{\mathrm{opt}}) - \bar P^\star\big)/{\bar P^\star}$, respectively, and ii) the peak power improvement of the \textsf{Proposed} method over the \textsf{AVG} method, measured by the relative reduction $\big(\bar P(\mathbf{h}_0^{\mathrm{avg}}) - \bar P(\mathbf{h}_0^{\mathrm{opt}})\big)/{\bar P(\mathbf{h}_0^{\mathrm{avg}})}$. The \textsf{AVG} method, which relies on a predetermined, non-optimized base set $\mathbf h_0^{\mathrm{avg}}$, exhibits a significant gap compared to the best-case benchmark, with its median reaching 30\%. In contrast, our \textsf{Proposed} method, leveraging an optimized base set, reduces the median gap to a mere 7\%.  The resulting performance enhancement is significant, as the third boxplot shows a relative improvement of the \textsf{Proposed} method over the \textsf{AVG} benchmark consistently being 17-20 percentage points.  These results validate that our base set optimization unlocks substantial additional flexibility from the EV fleet, enabling more effective grid services like peak shaving. 
\cg{Note our proposed method performs federated base set optimization and flexibility aggregation by only exploiting the inherently decomposable structure of the objective function and its gradient. As a result, it achieves a zero performance gap when compared to the centralized solution of the base set optimization problem \eqref{eqn_centralized_2}.}

%
%
%

\begin{figure}[ht]
\vspace{-3mm}
\centering

\subfloat{\includegraphics[width=0.9\columnwidth]{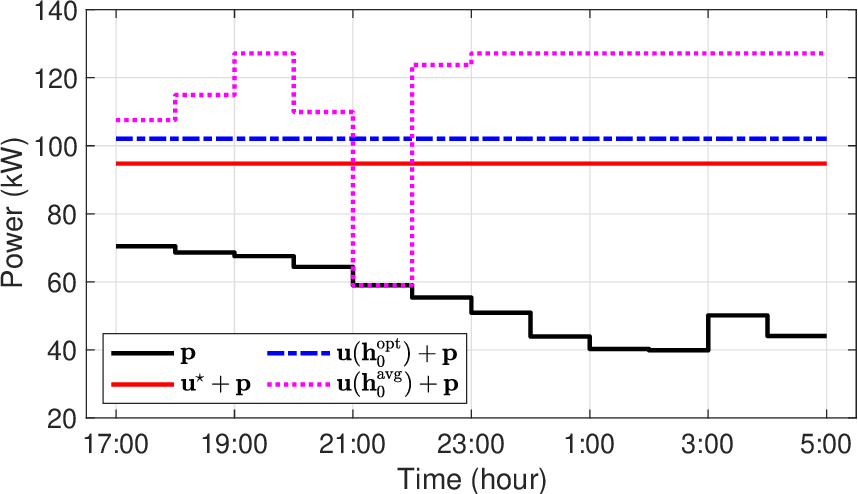}}
\vspace{-3mm}
\caption{Total power profiles for peak power minimization using historical load data from a commercial building on June 16, 2018, obtained from the ComStock dataset.}
\label{fig:profiles_peakPowerMin}
\vspace{-2mm}
\end{figure}

Fig.~\ref{fig:profiles_peakPowerMin} provides a qualitative comparison of the resulting total load profiles ($\mathbf{u}+\mathbf p$) for a representative load instance. The performance of our \textsf{Proposed} method in mitigating peak demand is immediately apparent. The total load curve corresponding to our method $\big(\mathbf{u}(\mathbf{h}_0^\mathrm{opt})+\mathbf p\big)$ exhibits a remarkable resemblance to the best-case benchmark $(\mathbf{u}^\star+\mathbf p)$. Both profiles demonstrate effective ``valley-filling" by strategically scheduling EV charging to flatten the original load curve, resulting in a near-constant power draw and a significantly reduced peak. In contrast, the profile resulting from the \textsf{AVG} benchmark $\big(\mathbf{u}(\mathbf{h}_0^\mathrm{avg})+\mathbf p\big)$ fails to effectively shave the peak. This erratic behavior, stemming directly from its sub-optimal characterization of the EV fleet's flexibility, leads to a substantially higher peak load. For example, our method achieves a 19.72\% reduction in peak power compared to this benchmark. These results further confirm the performance enhancement contributed by base set optimization.


\subsubsection{Electricity cost minimization} 
This task aims to re-schedule the charging of the EV fleet to minimize the total electricity cost, which can be modeled as
\begin{equation}\label{eq:cost_min}
 \abovedisplayskip=3pt
 \belowdisplayskip=3pt
    \min_{\mathbf u} \ (\bm \pi^\top \mathbf u)\Delta t,  \ \ \text{s.t. } \   \mathbf u\in \widetilde{\mathcal U},
\end{equation}
where $\bm \pi \in \mathbb R^T$ (\$/kWh) is the \cg{wholesale electricity price.} 
For the price profiles, we use historical zonal locational-based marginal prices (LMPs) from the day-ahead (DA) market operated by NYISO \cite{nyiso2025} (363 instances throughout the year 2024). 
Similar to the previous task, we also solve the following \emph{centralized optimization} problem as a best-case benchmark: \begin{equation}\label{eq:cost_min_centralize}
\abovedisplayskip=3pt
\belowdisplayskip=3pt
    \Pi^\star = \min_{\mathbf u_i, \forall i \in \mathcal N} \ \bigg(\! \bm \pi^\top \sum_{i\in \mathcal N}\mathbf u_i \!\bigg)\Delta t \ \ \text{s.t. }\  \mathbf u_i \in \mathcal U_i, \, \forall i \in \mathcal N.
\end{equation}

\begin{figure}[ht]
	\vspace{-3mm}
	\centering	{\includegraphics[width=0.9\columnwidth]{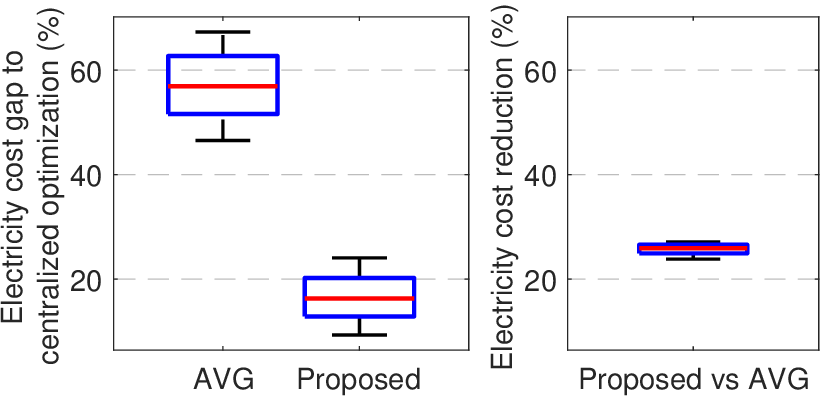}}
	\vspace{-3mm}
	\caption{Distributions of the electricity cost gaps and improvements for the electricity cost minimization problem~\eqref{eq:cost_min} under different methods.}
	\vspace{-2mm}
	\label{gapDistributions_2}
\end{figure}

Fig.~\ref{gapDistributions_2} illustrates i) the relative electricity cost gaps of the \textsf{AVG} and \textsf{Proposed} methods against the best-case benchmark, computed as $\big(\Pi(\mathbf{h}_0^{\mathrm{avg}}) - \Pi^\star\big)/{\Pi^\star}$ and $\big(\Pi(\mathbf{h}_0^{\mathrm{opt}}) - \Pi^\star\big)/{\Pi^\star}$, respectively, and ii) the electricity cost improvement of the \textsf{Proposed} method over the \textsf{AVG} method, measured by the relative reduction $\big(\Pi(\mathbf{h}_0^{\mathrm{avg}}) - \Pi(\mathbf{h}_0^{\mathrm{opt}})\big)/{\Pi(\mathbf{h}_0^{\mathrm{avg}})}$. While both methods incur some level of optimality gaps compared to the best-case benchmark, our \textsf{Proposed} method consistently demonstrates superior performance. Employing the optimized base-set parameter $\mathbf{h}_0^{\mathrm{opt}}$ markedly narrows the gap. As highlighted by the third boxplot, this translates into a substantial relative cost savings, with our approach consistently outperforming the \textsf{AVG} benchmark by a significant margin. These results provide strong evidence that a more accurate flexibility characterization, as enabled by our method, directly translates into economic value.


\begin{figure}[ht]
\centering
\subfloat{\includegraphics[width=0.9\columnwidth]{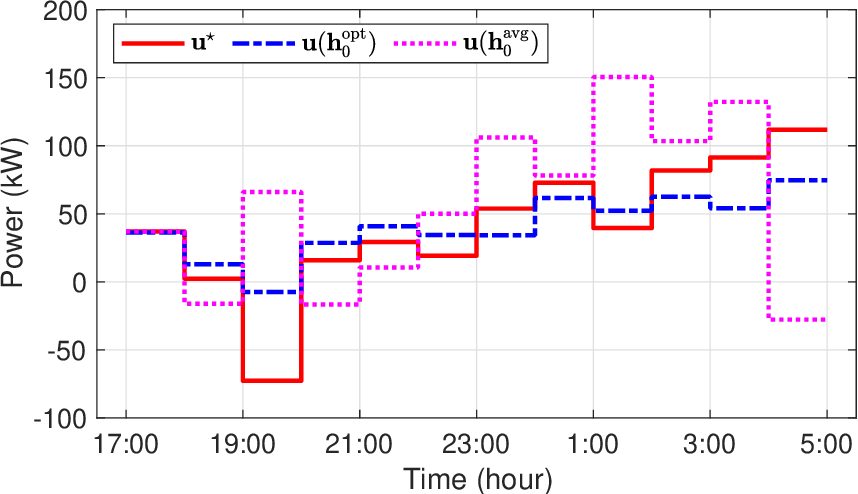}}

\vspace{2mm}

\  \!   
\subfloat{\includegraphics[width=0.875\columnwidth]{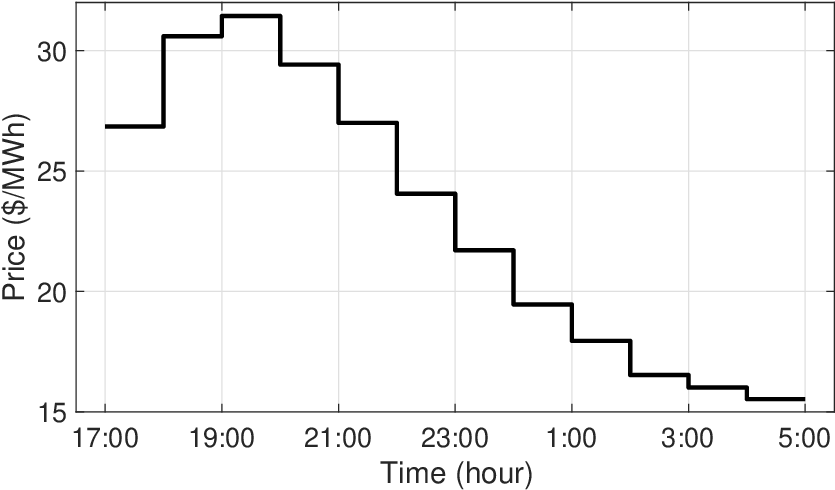}}
\vspace{-3mm}
\caption{Aggregate power profiles for electricity cost minimization using data obtained from NYISO Central Zone DA LBMPs on June 8, 2024.}
 \vspace{-4mm}
\label{fig:profiles_costMin}
\end{figure}

Fig.~\ref{fig:profiles_costMin} illustrates how an optimized base set enables more cost-effective charging behavior. For a specific electricity price profile, shown in the bottom panel, we compare the resulting aggregate charging dispatch profiles, shown in the top panel. The dispatch profile from our \textsf{Proposed} method, $\mathbf{u}(\mathbf{h}_0^\mathrm{opt})$, demonstrates a remarkable alignment with the best-case benchmark, $\mathbf{u}^\star$. It effectively performs temporal arbitrage: charging aggressively during low-price periods (e.g., 4:00-5:00) and minimizing power consumption or even discharging during high-price intervals (e.g., 19:00-20:00). This behavior confirms that our optimized base set accurately captures the temporal dependencies required for cost-efficient energy shifting. \cg{In contrast, the \textsf{AVG} benchmark employs an unoptimized, average-based parameter, $\mathbf{h}_0^\mathrm{avg}$. This parameter defines the shape of the aggregate flexibility set $\widetilde{\mathcal U}(\mathbf{h}_0^\mathrm{avg})$ via \eqref{eq:agg_flex_set}, and may exclude a subset of the \emph{true} aggregate flexibility set $\mathcal U$ that contains cost-effective dispatch strategies. Therefore, the resulting power profile, $\mathbf{u}(\mathbf{h}_0^\mathrm{avg})$, exhibits economically counter-intuitive behavior.}
For instance, it schedules significant charging at 19:00, an hour with a relatively high price. This misalignment with price signals directly leads to higher overall costs. 
\cg{As a result, the optimized base set parameter $\mathbf h_0^\mathrm{opt}$ achieves a 27.08\% reduction in electricity cost compared to the average-based parameter $\mathbf h_0^\mathrm{avg}$.}




\section{Conclusion}\label{sec:conclusion}

This paper proposes a federated framework for demand flexibility aggregation. In contrast to existing decentralized geometric methods that rely on a pre-determined, heuristic base set, our framework establishes a true federated process by enabling the collaborative optimization of this shared template. To achieve this, we first formulate the base set optimization as a non-convex bilevel program and then convert it into a single-level, unconstrained learning task. Leveraging the decomposable structure of this task's objective gradient, we design a federated gradient-based algorithm for its efficient, decentralized solution. By integrating this base set optimization with existing protocols for federated aggregation and disaggregation, we establish a complete and practical workflow for flexibility aggregation where only non-sensitive information is exchanged. Numerical results validate the significant benefits of our framework. The ability to collaboratively optimize a base set tailored to the specific resource population allows our approach to unlock substantially more flexibility than methods using a static base set. This enhanced characterization directly translates to superior performance in downstream grid service applications.

For future work, we plan to extend the framework to explicitly account for uncertainties in DSR behavior. Furthermore, incorporating detailed power network constraints represents a promising direction to further enhance the practical applicability of our adaptive aggregation methodology.

\appendices
\setcounter{table}{0}  
\section{Gradient Derivation \grn{for} Projection Problem} \label{app:projection}
\grn{This appendix derives the gradient of the projection mapping $\widetilde{\mathbf h}_0(\mathbf h_0)$, the solution of the projection problem \eqref{eqn_projection}, with respect to the base set parameter $\mathbf h_0$. The Lagrangian function of problem \eqref{eqn_projection} is}
\begin{equation}
\abovedisplayskip=2pt
\belowdisplayskip=2pt
    \mathcal L(\mathbf h_0^\prime, \mathbf u, \bm \lambda) := ||\mathbf h_0 - \mathbf h_0^\prime||_2^2 + \bm \lambda^\top(\mathbf H\mathbf u + \epsilon\mathbf 1 - \mathbf h_0^\prime),
\end{equation}where $\bm \lambda$ is the dual variable corresponding to the constraint. Let the optimal \jq{primal-dual} solution be $({\mathbf h_0^\prime}^{\! \star}, \mathbf u^\star, \bm \lambda^\star)$. The KKT conditions are given by
\begin{subequations}
    \abovedisplayskip=1pt
    \belowdisplayskip=1pt
    \begin{align}
        & \nabla_{\mathbf{h}_0^\prime} \mathcal{L}=-2(\mathbf{h}_0 - \mathbf{h}_0'^\star) - \bm{\lambda}^\star = \mathbf{0}, \\
        & \nabla_{\mathbf{u}} \mathcal{L}=\mathbf{H}^T \bm{\lambda}^\star = \mathbf{0}, \\
        & \mathbf{g}^\star := \mathbf{H}\mathbf{u}^\star + \epsilon\mathbf 1 - \mathbf{h}_0'^\star \leq \mathbf{0}, \\
        & \bm{\lambda}^\star \geq \mathbf{0}, \quad \bm{\lambda}^\star \circ \mathbf{g}^\star = \mathbf{0},
    \end{align}
\end{subequations}\grn{where $\circ$ denotes the Hadamard (element-wise) product.} We now define the KKT residual system as a vector-valued function $\mathbf F(\mathbf y, \mathbf h_0)$ \grn{where $\mathbf{y} := \big[{\mathbf{h}_0'^{\star}}^{\!\top}, {\mathbf{u}^{\star}}^{\!\top}, {\bm{\lambda}^{\star}}^{\!\top}\big]^\top \!\in\! \mathbb R^{9T}$}. 
We stack the relevant KKT equations and obtain \begin{equation}
\abovedisplayskip=2pt
\belowdisplayskip=2pt
    \mathbf{F}(\mathbf y, \mathbf{h}_0) = 
\begin{bmatrix}
2(\mathbf{h}_0 - \mathbf{h}_0'^\star) + \bm{\lambda}^\star \\
\mathbf{H}^\top \bm{\lambda}^\star \\
\bm{\lambda}^\star \circ \mathbf{g}^\star
\end{bmatrix} = \mathbf{0}.
\end{equation}
Then, the Jacobian of the function $\mathbf F$ with respect to $\mathbf y$ is \begin{equation}
\abovedisplayskip=2pt
\belowdisplayskip=2pt
    \mathbf J_\mathbf y := \begin{bmatrix}
-2\mathbf{I}_{4T} & \mathbf{0}_{4T \times T} & \mathbf{I}_{4T}\\
\mathbf{0}_{T \times 4T} & \mathbf{0}_{T \times T} & \mathbf{H}^\top \\
-\operatorname{diag}(\bm{\lambda}^\star) & \operatorname{diag}(\bm{\lambda}^\star)\cdot\mathbf{H} & \operatorname{diag}(\mathbf{g}^\star)
\end{bmatrix}.
\end{equation}Finally, we can calculate the gradient \begin{equation}
\abovedisplayskip=2pt
\belowdisplayskip=2pt
    \frac{\partial \mathbf y}{\partial \mathbf h_0} = - \mathbf J^{-1}_\mathbf y \cdot \frac{\partial \mathbf F}{\partial \mathbf h_0} = - \mathbf J^{-1}_\mathbf y \cdot \begin{bmatrix}
2\mathbf{I}_{4T} \\
\mathbf{0}_{T \times 4T} \\
\mathbf{0}_{T \times 4T}
\end{bmatrix}.
\end{equation}The top $4T$ rows of this matrix correspond to $\nabla_{\mathbf h_0} \mathbf h_0^\prime$, i.e., the desired gradient of the projected parameter with respect to the original base-set vector.

\vspace{-1mm}
\section{Gradient Derivation \grn{for} Set Containment Problem} \label{app:setContain}
\grn{This appendix derives the gradient of the optimal affine-transformation matrix $\bm \Gamma_i^\star$, the solution to the set containment problem \eqref{eq:opt_EV}, with respect to the base set parameter $\mathbf h_0$}. We suppress the subscript $i$ for readability. We first introduce Lagrange multipliers $\mathbf Y \in \mathbb R^{4T\times T}$, $\bm \mu \in \mathbb R^{4T}$ and $\bm \nu \in \mathbb R^{4T\times 4T}$ for constraints $\bm \Lambda \mathbf H = \mathbf H \bm \Gamma$, $\bm \Lambda  {\mathbf h}_0 \leq \mathbf h - \mathbf H \bm \gamma$ and $\bm \Lambda \geq \bm 0$, respectively. Using the Frobenius inner product $\langle \mathbf A, \mathbf B \rangle := \mathrm{Tr}(\mathbf A^\top\mathbf B)$, the Lagrangian of problem \eqref{eq:opt_EV} is \begin{equation*}
\abovedisplayskip=2pt
\belowdisplayskip=2pt
\begin{aligned}
    \mathcal L(\bm \gamma, \bm \Gamma, \bm \Lambda; \mathbf Y, \bm \mu, \bm \nu)= & -\mathrm{Tr}(\bm \Gamma) + \langle \mathbf Y,  \bm \Lambda \mathbf H - \mathbf H \bm \Gamma\rangle  \\
    & + \bm \mu^\top(\bm \Lambda  {\mathbf h}_0 + \mathbf H \bm \gamma - \mathbf h) - \langle \bm \nu, \bm\Lambda \rangle.
\end{aligned}
\end{equation*}Let $(\bm \gamma^\star , \bm \Gamma^\star , \bm \Lambda^\star; \mathbf Y^\star, \bm \mu^\star, \bm \nu^\star)$ be the optimal \grn{primal-dual} solution. The KKT conditions are given by
\begin{subequations}
\abovedisplayskip=3pt
\belowdisplayskip=3pt
\begin{align}
	& \nabla_{\bm \gamma} \mathcal{L}= \mathbf H^\top \bm \mu^\star = \mathbf{0}, \quad\! \nabla_{\bm \Gamma} \mathcal{L}= -\mathbf I_T - \mathbf H^\top \mathbf Y^\star = \mathbf{0},  \\
	& \nabla_{\bm \Lambda} \mathcal{L} = \mathbf Y^\star\mathbf H^\top + \bm \mu^\star {\mathbf h}_0^\top - \bm \nu^\star = \mathbf{0}, \\
	&  \bm \Lambda^\star \geq \bm 0, \quad  \bm \Lambda^\star \mathbf H - \mathbf H \bm \Gamma^\star = \mathbf 0, \\
	& \mathbf g^\star =  \bm \Lambda^\star  {\mathbf h}_0 + \mathbf H \bm \gamma^\star - \mathbf h \le \mathbf 0, \\
	& \bm \mu^\star \geq \mathbf{0},\quad \bm \nu^\star \ge \mathbf 0, \quad \bm \mu^\star \circ \mathbf g^\star = \mathbf 0, \quad \bm{\nu}^\star \circ  \bm \Lambda^\star = \mathbf{0},
\end{align}
\end{subequations}\grn{where $\circ$ denotes the Hadamard (element-wise) product.} We now define the KKT residual system as a vector-valued function \grn{$\mathbf G(\mathbf z, \mathbf h_0)$ where}
\begin{equation*}
\abovedisplayskip=2pt
\belowdisplayskip=2pt
    \grn{\mathbf z} := [{\bm \gamma^\star}^\top\!, \mathrm{vec}(\bm \Gamma^\star)^\top\!, \mathrm{vec}(\bm \Lambda^\star)^\top\!, \mathrm{vec}(\mathbf Y^\star)^\top, {\bm \mu^\star}^\top\!, \mathrm{vec}(\bm \nu^\star)^\top]^\top,
\end{equation*}
and $\mathrm{vec(\cdot)}$ denotes column-major vectorization (stacking the columns of a matrix into a single vector).
We stack the relevant KKT equations and obtain \begin{equation}\label{eq:ini_F}
\abovedisplayskip=0pt
\belowdisplayskip=2pt
    \mathbf{G}(\mathbf z, \mathbf{h}_0) = 
\begin{bmatrix}
		\mathbf H^\top \bm \mu^\star \\
		-\mathbf I_T - \mathbf H^\top \mathbf Y^\star \\
		\mathbf Y^\star\mathbf H^\top + \bm \mu^\star {\mathbf h}_0^\top - \bm \nu^\star \\
		\bm \Lambda^\star \mathbf H - \mathbf H \bm \Gamma^\star \\
		\bm \mu^\star \circ \mathbf g^\star \\
		\bm{\nu}^\star \circ  \bm \Lambda^\star \\
	\end{bmatrix}  = \mathbf{0}.
\end{equation}Applying the identities $\mathrm{vec}(\mathbf A \circ \mathbf B) = \mathrm{diag}(\mathrm{vec}(\mathbf A))\, \mathrm{vec}(\mathbf B)$, and $\mathrm{vec}(\mathbf A\mathbf B\mathbf C) = (\mathbf C^\top \!\otimes \mathbf A)\, \mathrm{vec}(\mathbf B)$ \grn{with $\otimes$ being the Kronecker product}, the vector-valued function $\mathbf G(\mathbf z, \mathbf h_0)$ becomes
\begin{equation*}
\abovedisplayskip=2pt
\belowdisplayskip=2pt
	\begin{aligned}
	    \mathbf G(\mathbf z, \mathbf h_0)& \!=\!\!\begin{bmatrix}
		\mathbf H^\top \bm \mu^\star \\
		\mathrm{vec}(-\mathbf I_T) - (\mathbf I_T \otimes \mathbf H^\top)\, \mathrm{vec}(\mathbf Y^\star) \\
		(\mathbf H \otimes \mathbf I_{4T})\, \mathrm{vec}(\mathbf Y^\star) + (\mathbf h_0 \otimes \mathbf I_{4T})\, \bm \mu^\star \!-\! \mathrm{vec}(\bm \nu^\star) \\
		(\mathbf H^\top \otimes \mathbf I_{4T})\, \mathrm{vec}(\bm \Lambda^\star) - (\mathbf I_T \otimes \mathbf H)\, \mathrm{vec}(\bm \Gamma^\star) \\
		\mathrm{diag}(\bm \mu^\star)\, \mathbf g^\star \\
		\mathrm{diag}(\mathrm{vec}(\bm{\nu}^\star))\,  \mathrm{vec}(\bm \Lambda^\star) \\
	\end{bmatrix}\!\!.
	\end{aligned}
\end{equation*}
Then, the Jacobian of the function $\mathbf G$ with respect to $\mathbf z$ is $\mathbf J_\mathbf z := \left[\nabla_{\!\bm \gamma^\star}\mathbf G, \nabla_{\!\mathrm{vec}(\bm \Gamma^\star\!)}\mathbf G,\nabla_{\!\mathrm{vec}(\bm \Lambda^\star\!)}\mathbf G,\nabla_{\!\mathrm{vec}(\mathbf Y^\star\!)}\mathbf G,\nabla_{\!\bm \mu^\star}\mathbf G,\nabla_{\!\mathrm{vec}(\bm \nu^\star\!)}\mathbf G\right]$, which is sparse with non-zero blocks being \begin{subequations}
    \abovedisplayskip=1pt
    \belowdisplayskip=2pt
    \begin{align*}
        & \mathbf J_\mathbf z[1, 5] = \mathbf H^\top, & \mathbf J_\mathbf z[2, 4] = - (\mathbf I_T \otimes \mathbf H^\top), \\
        & \mathbf J_\mathbf z[3, 4] = \mathbf H \otimes \mathbf I_{4T}, & \mathbf J_\mathbf z[3, 5] = \mathbf h_0 \otimes \mathbf I_{4T}, \\
        & \mathbf J_\mathbf z[3, 6] = -\mathbf I_{4T}, & \mathbf J_\mathbf z[4, 2] = - (\mathbf I_T \otimes \mathbf H), \\
        &\mathbf J_\mathbf z[4,3] = \mathbf H^\top \!\otimes \mathbf I_{4T}, & \mathbf J_\mathbf z[5, 1] = \mathrm{diag}(\bm \mu^\star)\mathbf H,\\
        &\mathbf J_\mathbf z[5, 3] = \mathrm{diag}(\bm \mu^\star) (\mathbf h_0^\top\! \otimes \mathbf I_{4T}), & \mathbf J_\mathbf z[5, 5] =\mathrm{diag}(\mathbf g^\star), \\
        &\mathbf J_\mathbf z[6, 3] = \mathrm{diag}(\mathrm{vec}(\bm{\nu}^\star)), & \mathbf J_\mathbf z[6,6] = \mathrm{diag}(\mathrm{vec}(\bm{\Lambda}^\star)).
    \end{align*}
\end{subequations}
Finally, we can calculate the gradient \begin{equation}
\abovedisplayskip=0pt
\belowdisplayskip=2pt
	\frac{\partial \mathbf z}{\partial \mathbf h_0} = -\mathbf J_{\mathbf z}^{-1} \, \frac{\partial \mathbf G}{\partial \mathbf h_0} = -\mathbf J_{\mathbf z}^{-1} \begin{bmatrix}
		\mathbf 0 \\
		\mathbf 0 \\
		\mathbf I_{4T} \otimes \bm \mu^\star \\
		\mathbf 0 \\
		\mathrm{diag}(\bm \mu^\star) \, \bm \Lambda^\star \\
		\mathbf 0 \\
	\end{bmatrix}.
\end{equation}By extracting the block of this gradient which corresponds to $\mathrm{vec}(\bm \Gamma^\star)$, i.e., rows $T+1$ through $T+T^2$, we can obtain $\nabla_{\mathbf h_0}\mathrm{vec}(\bm\Gamma^\star) \in\mathbb R^{T^2 \times 4T}$. This can be reshaped into a $T\times T\times 4T$ tensor corresponding to $\nabla_{\mathbf h_0}\bm \Gamma^\star$.

\section{Nested Integral Representation of Base-Set Volume and Its Gradient}\label{app:vol}
For a given $\mathbf h_0 \!:=\! [\overline{\mathbf x}, -\underline{\mathbf x}, \overline{\mathbf u}, -\underline{\mathbf u}]$, it holds that for any $t\in \mathcal T$, \begin{equation*}
\abovedisplayskip=3pt
\belowdisplayskip=3pt
    x(t) - x(t-1) \in [\underline{u}(t) \!\cdot\! \Delta t,\ \overline{u}(t)\!\cdot\! \Delta t], \ x(t) \in [\underline{x}(t), \ \overline{x}(t)],
\end{equation*}where $x(0) := 0$. To compute the volume defined in \eqref{eq:vol_def}, we first define the admissible interval $[a_t, b_t]$ for dimension $t$ as
\begin{subequations}
\abovedisplayskip=0pt
\belowdisplayskip=0pt
    \begin{align}
        & a_t := a_t(x(t\!-\!1)) = \max(x(t\!-\!1) + \underline{u}(t)\!\cdot\! \Delta t, \ \underline{x}(t)),\\
        & b_t := b_t(x(t\!-\!1)) = \min(x(t\!-\!1) + \overline{u}(t)\!\cdot\! \Delta t, \ \overline{x}(t)).
    \end{align}
\end{subequations}
Then, the base set volume can be written as a $T$-fold integral:
\begin{equation}\label{eq:vol_expression}
\abovedisplayskip=2pt
\belowdisplayskip=2pt
\begin{aligned}
    \mathrm{vol}(\mathcal U_0(\mathbf h_0)) := \int_{\mathbf x\in \mathbb R^T}\,  \prod_{t=1}^{T} \mathbb{I}\big\{x(t)\in[a_t, b_t]\big\}\, \mathrm{d}\mathbf x. 
\end{aligned}
\end{equation}
To evaluate this integral recursively, we define $R_t(x(t))$ as the accumulated feasible mass after $t$ steps ending at state $x(t)$:
\begin{equation}\label{recursiveF}
\abovedisplayskip=2pt
\belowdisplayskip=2pt
    R_t(x(t)) := \int_{x(t-1)}\! R_{t-1}(x(t-1))\cdot \mathbb I\big\{x(t)\!\in\![a_t, b_t]\big\}\, \mathrm{d}x(t-1),
\end{equation}where $R_0(x(0)) := \delta(x(0))$ with $\delta(\cdot)$ being the \emph{Dirac delta function}. Under this definition, \eqref{eq:vol_expression} can be rewritten as
\begin{equation}
\abovedisplayskip=2pt
\belowdisplayskip=2pt
\begin{aligned}
    \mathrm{vol}(\mathcal U_0(\mathbf h_0)) = \int_{x(T)} R_T(x(T)) \, \mathrm{d}x(T).
\end{aligned}
\end{equation}

To differentiate the base-set volume with respect to $\mathbf h_0$, \grn{we take partial derivatives coordinate-wise. Let $\mathbf h_0 := [\overline{\mathbf x}, -\underline{\mathbf x}, \overline{\mathbf u}, -\underline{\mathbf u}]$ with $\overline{\mathbf x}:=[\overline x(t)]_{t\in\mathcal T}$, $\underline{\mathbf x}:=[\underline x(t)]_{t\in\mathcal T}$, $\overline{\mathbf u}:=[\overline u(t)]_{t\in\mathcal T}$, and $\underline{\mathbf u}:=[\underline u(t)]_{t\in\mathcal T}$. Since the gradient with respect to each bound in $\mathbf h_0$ is different, we use $\theta(t)$ as a general notation to refer to one of $\overline{x}(t)$, $\underline{x}(t)$, $\overline{u}(t)$ and $\underline{u}(t)$, and compute $\nabla_{\theta(t)}\, \mathrm{vol}(\mathcal U_0(\mathbf h_0))$ using the nested representation.}
Specifically, according to the Leibniz integral rule, we have
\begin{equation}\label{eq:gradients}
\abovedisplayskip=4pt
\belowdisplayskip=4pt
    \nabla_{\theta(t)} \,\mathrm{vol}(\mathcal U_0(\mathbf h_0)) = \int_{x(T)} \nabla_{\theta(t)}\, R_T(x(T)) \ \mathrm{d}x(T).
\end{equation}
We expand $R_T(x(T))$ recursively via \eqref{recursiveF} until dimension $t$, noting that $\theta(t)$ appears only through the bounds ($a_t$ or $b_t$) of the $t$-th indicator, while all bounds from dimension 1 to $t\!-\!1$ are independent of $\theta(t)$. When $\theta(t) = \overline{x}(t)$, we finally obtain
\begin{subequations}
\abovedisplayskip=2pt
\belowdisplayskip=2pt
    \begin{align*}
        & \nabla_{\theta(t)}\, R_T(x(T)) \!=\! \int \underbrace{\prod_{\tau = t+1}^{T}\!\! \mathbb I\big\{x(\tau) \!\in\! [a_\tau, b_\tau]\big\} \!\cdot\! \delta(x(t) \!-\! \overline{x}(t))}_{\text{independent of}\ x(t-1)} \\
        & \!\!\!\!\underbrace{R_{t\!-\!1}(x(t\!\!-\!\!1)) \!\cdot\! \mathbb I\big\{\!a_t \!\le\! b_t\!\big\} \!\cdot\! \mathbb I\big\{\overline{x}(t) \!<\! x(t\!\!-\!\!1) \!+\! \overline{u}(t)\!\cdot\!\Delta t\big\}}_{\text{dependent only on}\ x(t-1)} \mathrm{d}\mathbf x(t\!\!-\!\!1 \!:\! T) \\
        & = S(R_{t-1})\!\int \delta(x(t) \!-\! \overline{x}(t)) \cdot\!\! \prod_{\tau = t+1}^{T}\!\! \mathbb I\big\{x(\tau) \!\in\! [a_\tau, b_\tau]\big\}\, \mathrm{d}\mathbf x(t \!:\! T),  
    \end{align*}
\end{subequations}
\grn{where $\mathrm{d}\mathbf x(i\!:\!j) \!:=\! \mathrm{d}x(i)\, \mathrm{d}x(i\!+\!1) \cdots \mathrm{d}x(j)$, and $S(R_{t-1})$ denotes the result of the integration over $x(t-1)$, depending on the function $R_{t-1}$. This derivative shares a similar nested structure as \eqref{eq:vol_expression}, and can be recursively evaluated given the sequence of functions $R_t(x(t))$ for all $t\in\mathcal T$ calculated via \eqref{recursiveF}. Similar derivations follow when $\theta(t)\in\{\underline{x}(t),\, \overline{u}(t),\, \underline{u}(t)\}$.}

\section{Numerical Recursive Integration Scheme}\label{app:mc_integration}
This appendix presents a numerical recursive integration scheme that yields computationally efficient estimates of the base-set volume $\mathrm{vol}(\mathcal U_0(\mathbf h_0))$ and its gradient $\nabla_{\mathbf h_0}\mathrm{vol}(\mathcal U(\mathbf h_0))$. \grn{Appendix \ref{app:vol} presents a nested representation of the base-set volume and its gradient, which enables a dimension-by-dimension computation. Leveraging this structure, at each recursion level, we estimate the one-dimensional integral with a Riemann-sum approximation on a \emph{fixed and bounded} interval instead of the \emph{state-dependent} interval described in Appendix \ref{app:vol}, and proceed dimension by dimension.}

For each $t\in \mathcal T$, we discretize $x(t)$ on the following fixed interval with step size $\Delta x$:
\begin{equation}\label{eq:samplingInterval}
\abovedisplayskip=3pt
\belowdisplayskip=3pt
    \mathcal{I}_t := \left[\!\max\!\bigg(\underline x(t), \sum_i^t\underline{u}(t)\!\cdot\!\Delta t \bigg), \ \min\!\bigg(\overline x(t), \sum_i^t\overline{u}(t)\!\cdot\!\Delta t \bigg)\!\right]\!.
\end{equation}\grn{This interval is the tightest range that is independent of past states but guaranteed to contain all feasible $x(t)$, discretizing over which keeps the recursion simple (no dependence on the state-dependent bounds $a_t$ and $b_t$ introduced in Appendix \ref{app:vol}) while avoiding evaluating outside the physically attainable range. With step size $\Delta x$, discretizing interval $\mathcal I_t$ yields $K_t := |\mathcal{I}_t|/\Delta x$ points. Algorithm \ref{alg:MC} then builds a recursive estimator as follows: the input consists of $K_1$ points $x^{(k)}\!(1)$ together with the corresponding $R_1\big(x^{(k)}\!(1)\big) := \mathbb I\big\{x^{(k)}\!(1) \in [a_1(0), b_1(0)]\big\}$. For each $t>1$, discretize $\mathcal I_t$ and obtain $K_t$ points, and for each point $x^{(k)}\!(t)$, approximate $R_t\big(x^{(k)}\!(t)\big)$ by the following Riemann-sum discretization of \eqref{recursiveF}:
\begin{equation}\label{eq:F_approx}
\abovedisplayskip=3pt
\belowdisplayskip=3pt
    \begin{aligned}
        \widehat{R}_t(x^{(k)}\!(t)) \approx &\ \Delta x \cdot \sum_{k=1}^{K_{t-1}} R_{t-1}(x^{(k)}\!(t\!-\!1)) \\
        &\quad\mathbb I\Big\{x^{(k)}\!(t) \!\in\! \big[a_t\big(x^{(k)}\!(t\!-\!1)\big), b_t\big(x^{(k)}\!(t\!-\!1)\big)\big]\Big\},
    \end{aligned}
\end{equation}and cache the pair $\big(x^{(k)}\!(t), \, \widehat{R}_t\big(x^{(k)}\!(t)\big) \big)$ for the next recursion. Finally, we can estimate the volume via \begin{equation}\label{eq:volEst}
\abovedisplayskip=3pt
\belowdisplayskip=3pt
    \widehat{\mathrm{vol}}(\mathcal U_0(\mathbf h_0)) = \sum_{k=1}^{K_T}\widehat{R}_T(x^{(k)}\!(T)) \cdot\Delta x.
\end{equation}}

\vspace{-4mm}
\RestyleAlgo{ruled}
\setlength{\textfloatsep}{2pt}
\begin{algorithm}[hbt!]
	\caption{Numerical recursive integration}\label{alg:MC}
	\LinesNumbered
	\KwIn{Values of $x^{(k)}\!(1)$ and the corresponding values of $R_1(x^{(k)}\!(1))$.}

        \For{$t = 2, \dots, T$}{
		Discretize $\mathcal I_t$ and obtain $|K_t|$ values of $x^{(k)}\!(t)$. \\
            \For{$k = 1, \dots, K_t $}{
            Estimate $\widehat{R}_t\big(x^{(k)}\!(t)\big)$ via \eqref{eq:F_approx}.

            Store $\big(x^{(k)}\!(t), \, \widehat{R}_t\big(x^{(k)}\!(t)\big) \big)$ pairs.
            }
	}
	
	Estimate the volume via \eqref{eq:volEst}.
	
	\KwOut{Estimated $\widehat{\mathrm{vol}}(\mathcal U_0(\mathbf h_0))$, $\big(x^{(k)}\!(t), \, \widehat{R}_t\big(x^{(k)}\!(t)\big) \big)$ pairs for all $t\in\mathcal T$, $k \in \{1, \dots, K_t\}$. \\
	}
\end{algorithm}
\vspace{-4mm}

\grn{Since the gradient formulations share a similar nested structure to that of the volume (see Appendix \ref{app:vol}), we estimate each~component of the overall gradient via a similar dimension-by-dimension recursion, reusing the cached pairs in Algorithm 1.
}

\bibliographystyle{ieeetr}
\bibliography{manuscript_Extended.bib}

\begin{thebibliography}{10}

\bibitem{denholm2021challenges}
P.~Denholm, D.~J. Arent, S.~F. Baldwin, D.~E. Bilello, G.~L. Brinkman, J.~M.
  Cochran, W.~J. Cole, B.~Frew, V.~Gevorgian, J.~Heeter, {\em et~al.}, ``The
  challenges of achieving a 100\% renewable electricity system in the united
  states,'' {\em Joule}, vol.~5, no.~6, pp.~1331--1352, 2021.

\bibitem{10965352}
G.~Chen and J.~Qin, ``Neural risk limiting dispatch in power networks:
  Formulation and generalization guarantees,'' {\em IEEE Transactions on Power
  Systems}, pp.~1--13, 2025.

\bibitem{10700765}
G.~Chen, J.~Qin, and H.~Zhang, ``Model-free self-supervised learning for
  dispatching distributed energy resources,'' {\em IEEE Transactions on Smart
  Grid}, vol.~16, no.~2, pp.~1287--1300, 2025.

\bibitem{10681446}
X.~Huo and M.~Liu, ``On privacy preservation of distributed energy resource
  optimization in power distribution networks,'' {\em IEEE Transactions on
  Control of Network Systems}, vol.~12, no.~1, pp.~228--240, 2025.

\bibitem{9478223}
Z.~Su, Y.~Wang, T.~H. Luan, N.~Zhang, F.~Li, T.~Chen, and H.~Cao, ``Secure and
  efficient federated learning for smart grid with edge-cloud collaboration,''
  {\em IEEE Transactions on Industrial Informatics}, vol.~18, no.~2,
  pp.~1333--1344, 2022.

\bibitem{8063901}
F.~L. Müller, J.~Szabó, O.~Sundström, and J.~Lygeros, ``Aggregation and
  disaggregation of energetic flexibility from distributed energy resources,''
  {\em IEEE Transactions on Smart Grid}, vol.~10, no.~2, pp.~1205--1214, 2019.

\bibitem{7864461}
L.~Zhao, W.~Zhang, H.~Hao, and K.~Kalsi, ``A geometric approach to aggregate
  flexibility modeling of thermostatically controlled loads,'' {\em IEEE
  Transactions on Power Systems}, vol.~32, no.~6, pp.~4721--4731, 2017.

\bibitem{9029363}
S.~Sadraddini and R.~Tedrake, ``Linear encodings for polytope containment
  problems,'' in {\em 2019 IEEE 58th Conference on Decision and Control (CDC)},
  pp.~4367--4372, 2019.

\bibitem{7445245}
E.~Mayhorn, L.~Xie, and K.~Butler-Purry, ``Multi-time scale coordination of
  distributed energy resources in isolated power systems,'' {\em IEEE
  Transactions on Smart Grid}, vol.~8, no.~2, pp.~998--1005, 2017.

\bibitem{10073567}
Z.~Tan, Z.~Yan, H.~Zhong, and Q.~Xia, ``Non-iterative solution for coordinated
  optimal dispatch via equivalent projection—part i: Theory,'' {\em IEEE
  Transactions on Power Systems}, vol.~39, no.~1, pp.~890--898, 2024.

\bibitem{10878458}
J.~Thran, J.~Marecek, R.~N. Shorten, and T.~C. Green, ``Reserve provision from
  electric vehicles: Aggregate boundaries and stochastic model predictive
  control,'' {\em IEEE Transactions on Power Systems}, pp.~1--12, 2025.

\bibitem{11045802}
M.~Elsaadany, M.~R. Almassalkhi, and S.~H. Tindemans, ``Linear aggregate model
  for realizable dispatch of homogeneous energy storage,'' {\em IEEE Control
  Systems Letters}, vol.~9, pp.~1267--1272, 2025.

\bibitem{7463483}
H.~Zhang, Z.~Hu, Z.~Xu, and Y.~Song, ``Evaluation of achievable vehicle-to-grid
  capacity using aggregate pev model,'' {\em IEEE Transactions on Power
  Systems}, vol.~32, no.~1, pp.~784--794, 2017.

\bibitem{BRINKEL2023100297}
N.~Brinkel, L.~Visser, W.~{van Sark}, and T.~AlSkaif, ``A novel forecasting
  approach to schedule aggregated electric vehicle charging,'' {\em Energy and
  AI}, vol.~14, p.~100297, 2023.

\bibitem{9851563}
Y.~Wen, Z.~Hu, and L.~Liu, ``Aggregate temporally coupled power flexibility of
  ders considering distribution system security constraints,'' {\em IEEE
  Transactions on Power Systems}, vol.~38, no.~4, pp.~3884--3896, 2023.

\bibitem{10124221}
C.~Gu, J.~Wang, and L.~Wu, ``Distributed energy resource and energy storage
  investment for enhancing flexibility under a tso-dso coordination
  framework,'' {\em IEEE Transactions on Automation Science and Engineering},
  vol.~21, no.~3, pp.~2961--2973, 2024.

\bibitem{10347534}
Y.~Wen, Z.~Hu, J.~He, and Y.~Guo, ``Improved inner approximation for
  aggregating power flexibility in active distribution networks and its
  applications,'' {\em IEEE Transactions on Smart Grid}, vol.~15, no.~4,
  pp.~3653--3665, 2024.

\bibitem{9482808}
B.~Cui, A.~Zamzam, and A.~Bernstein, ``Network-cognizant time-coupled aggregate
  flexibility of distribution systems under uncertainties,'' in {\em 2021
  American Control Conference (ACC)}, pp.~4178--4183, 2021.

\bibitem{10286155}
D.~Yan, S.~Huang, and Y.~Chen, ``Real-time feedback based online aggregate ev
  power flexibility characterization,'' {\em IEEE Transactions on Sustainable
  Energy}, vol.~15, no.~1, pp.~658--673, 2024.

\bibitem{9383807}
X.~Chen and N.~Li, ``Leveraging two-stage adaptive robust optimization for
  power flexibility aggregation,'' {\em IEEE Transactions on Smart Grid},
  vol.~12, no.~5, pp.~3954--3965, 2021.

\bibitem{10490133}
F.~A. Taha, T.~Vincent, and E.~Bitar, ``An efficient method for quantifying the
  aggregate flexibility of plug-in electric vehicle populations,'' {\em IEEE
  Transactions on Smart Grid}, pp.~1--1, 2024.

\bibitem{10121812}
M.~Zhang, Y.~Xu, X.~Shi, and Q.~Guo, ``A fast polytope-based approach for
  aggregating large-scale electric vehicles in the joint market under
  uncertainty,'' {\em IEEE Transactions on Smart Grid}, vol.~15, no.~1,
  pp.~701--713, 2024.

\bibitem{9527324}
N.~Nazir and M.~Almassalkhi, ``Guaranteeing a physically realizable battery
  dispatch without charge-discharge complementarity constraints,'' {\em IEEE
  Transactions on Smart Grid}, vol.~14, no.~3, pp.~2473--2476, 2023.

\bibitem{10384282}
M.~Elsaadany and M.~R. Almassalkhi, ``Battery optimization for power systems:
  Feasibility and optimality,'' in {\em 2023 62nd IEEE Conference on Decision
  and Control (CDC)}, pp.~562--569, 2023.

\bibitem{nphard}
H.~Tiwary, ``On the hardness of computing intersection, union and minkowski sum
  of polytopes,'' {\em Discrete \& Computational Geometry}, vol.~40,
  pp.~469--479, 10 2008.

\bibitem{ab4575cf4ab642d89d2a88041b41675e}
J.~Nocedal and S.~Wright, ``Numerical optimization,'' {\em Springer Series in
  Operations Research and Financial Engineering}, pp.~1--664, 2006.

\bibitem{doi:10.1137/0217060}
M.~E. Dyer and A.~M. Frieze, ``On the complexity of computing the volume of a
  polyhedron,'' {\em SIAM Journal on Computing}, vol.~17, no.~5, pp.~967--974,
  1988.

\bibitem{BAROT201755}
S.~Barot and J.~A. Taylor, ``A concise, approximate representation of a
  collection of loads described by polytopes,'' {\em International Journal of
  Electrical Power \& Energy Systems}, vol.~84, pp.~55--63, 2017.

\bibitem{10.1145/3584182}
A.~Chalkis, I.~Z. Emiris, and V.~Fisikopoulos, ``A practical algorithm for
  volume estimation based on billiard trajectories and simulated annealing,''
  {\em ACM J. Exp. Algorithmics}, vol.~28, May 2023.

\bibitem{6572301}
M.~Wu, B.~Yin, A.~Vosoughi, C.~Studer, J.~R. Cavallaro, and C.~Dick,
  ``Approximate matrix inversion for high-throughput data detection in the
  large-scale mimo uplink,'' in {\em 2013 IEEE International Symposium on
  Circuits and Systems (ISCAS)}, pp.~2155--2158, 2013.

\bibitem{HiriartUrruty1996}
J.-B. Hiriart-Urruty and C.~Lemaréchal, {\em Convex Analysis and Minimization
  Algorithms I: Fundamentals}.
\newblock Berlin, Germany: Springer, 1996.

\bibitem{BallMilman1999}
K.~M. Ball and V.~D. Milman, eds., {\em Convex Geometric Analysis}.
\newblock Cambridge, UK: Cambridge University Press, 1999.

\bibitem{10057479}
J.~Mathias, A.~Bušić, and S.~Meyn, ``Load-level control design for demand
  dispatch with heterogeneous flexible loads,'' {\em IEEE Transactions on
  Control Systems Technology}, vol.~31, no.~4, pp.~1830--1843, 2023.

\bibitem{parker2023comstock}
A.~Parker, H.~Horsey, M.~Dahlhausen, M.~Praprost, C.~CaraDonna, A.~LeBar, and
  L.~Klun, ``{ComStock Reference Documentation: Version 1},'' Tech. Rep.
  NREL/TP-5500-83819, National Renewable Energy Laboratory, Golden, CO, 2023.

\bibitem{nyiso2025}
{NYISO}, ``{New York Independent System Operator}.'' [Online]. Available:
  \url{https://www.nyiso.com/energy-market-operational-data}.

\end{thebibliography}
\end{document}